\documentclass[10pt,a4paper,dvipsnames,usenames]{article}
\usepackage{amssymb}
\usepackage{amsmath}
\usepackage{amsthm}
\usepackage{latexsym}
\usepackage[dvips]{epsfig}
\usepackage{mathrsfs}
\usepackage{eufrak}
\usepackage{bm}
\usepackage{stmaryrd}
\usepackage{authblk}
\usepackage{slashed}
\usepackage{bigints}
\usepackage{centernot}

\usepackage[dvipsnames]{xcolor}
\usepackage{tikz}
\theoremstyle{plain}
\newtheorem{proposition}{Proposition}
\newtheorem{lemma}{Lemma}
\newtheorem{theorem}{Theorem}

\def\bma{{\bm a}}
\def\bmb{{\bm b}}
\def\bmc{{\bm c}}
\def\bmd{{\bm d}}
\def\bme{{\bm e}}
\def\bmf{{\bm f}}
\def\bmg{{\bm g}}
\def\bmh{{\bm h}}
\def\bmi{{\bm i}}
\def\bmj{{\bm j}}
\def\bmk{{\bm k}}
\def\bml{{\bm l}}
\def\bmn{{\bm n}}

\def\bmu{{\bm u}}
\def\bmv{{\bm v}}

\def\bmx{{\bm x}}

\def\bmz{{\bm z}}

\def\bmzero{{\bm 0}}
\def\bmone{{\bm 1}}
\def\bmtwo{{\bm 2}}
\def\bmthree{{\bm 3}}

\def\bmL{{\bm L}}

\def\bmalpha{{\bm \alpha}}
\def\bmbeta{{\bm \beta}}
\def\bmgamma{{\bm \gamma}}

\def\bmxi{{\bm \xi}}

\def\bmomega{{\bm \omega}}

\def\bmphi{{\bm \phi}}

\def\bmsigma{{\bm \sigma}}

\def\bmtau{{\bm \tau}}
\def\bmupsilon{{\bm \upsilon}}
\def\bmell{{\bm \ell}}

\def\bmGamma{{\bm \Gamma}}

\def\bmpartial{{\bm \partial}}
\def\bmnabla{{\bm \nabla}}

\def\bmell{{\bm \ell}}

\usepackage{mathtools}
\usepackage{xparse}
\makeatletter
\newcommand{\raisemath}[1]{\mathpalette{\raisem@th{#1}}}
\newcommand{\raisem@th}[3]{\raisebox{#1}{$#2#3$}}
\makeatother
\NewDocumentCommand{\newrbar}{O{0pt} O{0pt}}{
  \ensuremath{\mathrlap{\raisemath{#2}{\hspace*{#1}{\mathchar'26\mkern-9mu}}}r}}

\newcounter{mnotecount}

\newcommand{\mnotex}[1]
{\protect{\stepcounter{mnotecount}}$^{\mbox{\footnotesize $\bullet$\themnotecount}}$ 
\marginpar{
\raggedright\tiny\em
$\!\!\!\!\!\!\,\bullet$\themnotecount: #1} }

\newcounter{mnote}

\usepackage{color}
\usepackage{xcolor}
\usepackage[normalem]{ulem}
\usepackage{soul}
\usepackage{hyperref}

\begin{document}
\title{\textbf{Conformal geodesics and the evolution of spacetimes with positive Cosmological constant}}
 
\author[1]{ Marica Minucci\footnote{E-mail
    address:{\tt m.minucci@qmul.ac.uk}}}

\affil[1]{School of Mathematical Sciences, Queen Mary, University of London,
Mile End Road, London E1 4NS, United Kingdom.}

\maketitle

\begin{abstract}
This article provides a discussion on the construction of conformal Gaussian gauge systems to study the evolution of solutions to the Einstein field equations with positive Cosmological constant. This is done by means of a gauge based on the properties of conformal geodesics. The use of this gauge, combined with the extended conformal Einstein field equations, yields evolution equations in the form of a symmetric hyperbolic system for which standard Cauchy stability results can be employed. This strategy is used to study the global properties of de Sitter-like spacetimes with constant negative scalar curvature. It is then adapted to study the evolution of the Schwarzschild-de Sitter spacetime in the static region near the conformal boundary. This review is based on Class. Quantum Grav. {\bf 38} 145026 and Class. Quantum Grav. {\bf 40} 145005.
\end{abstract}

\section{Introduction}
One of the main open problems in mathematical relativity is that of
the non-linear stability of spacetimes. In 1986, Friedrich provided the first result concerning the global stability of the de Sitter spacetime and a semi-global stability result for the Minkowski spacetime \cite{Fri86b, Fri86c}. These results are obtained by using the conformal Einstein field equations to reformulate Cauchy problems which are global or semi-global in time into problems which are local in time.
This strategy allows to use results obtained for quasi-linear symmetric hyperbolic systems \cite{Kat75b, Kat75c} to prove the existence of solutions which are suitably close to known reference spacetimes. More results \cite{AlhMenVal17, Fri91, LubVal09, LubVal10, LubVal12, JourThaVal21, CFEBook} using the  \textit{conformal Einstein field equations} show that these equations are a
powerful tool for the analysis of the stability of spacetimes. They provide a system of field equations for geometric objects defined on a four-dimensional Lorentzian manifold $(\mathcal{M}, \bmg)$, the so-called
\textit{unphysical spacetime}, which is conformally related to a
spacetime $(\tilde{\mathcal{M}}, \tilde{\bmg})$, the so-called
\textit{physical spacetime}, satisfying the Einstein field
equations. The conformal Einstein field equations constitute a system of
differential conditions on the curvature tensors with respect to the
Levi-Civita connection of $\bmg$ and the conformal factor $\Xi$. 

\medskip
A problem one encounters when discussing the conformal structure of spacetimes by means of these equations is that of the gauge freedom. In the original formulation of the conformal Einstein field equations \cite{Fri84} the gauge is fixed by means of \emph{gauge source functions}. An alternative approach of gauge fixing is by exploiting the properties of a congruence of curves which are invariants of the conformal structure. These curves are known as \emph{conformal geodesics} and they have been originally introduced as a tool for the local analysis of the structure of conformally rescaled spacetimes \cite{FriSch87}.  Using this gauge allows to
define a \emph{conformal Gaussian gauge system} in which coordinates are propagated along conformal geodesics. To combine this gauge choice with the conformal Einstein
field equations it is necessary to make use of a more general version
of the latter, the \textit{extended conformal Einstein field
equations}. These equations contain a bigger gauge freedom being expressed using a \emph{Weyl connection}. This is a torsion-free connection which provides a transport equation along the conformal geodesics preserving conformally orthonormal frames and the causal nature of their vectors. 

\medskip
One of the advantages of the \emph{conformal Gaussian gauge system} is that it gives an a priori knowledge of the structure of the conformal boundary of the spacetime.
This aspect is used to obtain an alternative proof of the semi-global non-linear stability of the Minkowski spacetime and of the global non-linear stability of the de Sitter spacetime by L\"{u}bbe and Valiente Kroon \cite{LubVal09}. In \cite{MinVal21} the results obtained in \cite{Fri86b, LubVal09} are generalised to de Sitter-like spacetimes with compact spatial sections of negative scalar curvature. The existence and stability result follows from explicit calculations and the requirement that the data are close to de Sitter-like data. 
The success of this approach in the analysis of the global properties of asymptotically simple spacetimes leads to the question of whether a similar strategy can be used to study the evolution of black hole spacetimes. A first step in this direction is made in \cite{MinVal23} where certain aspects of the conformal structure of the sub-extremal Schwarzschild-de Sitter spacetime are analysed in order to adapt techniques from the asymptotically simple setting to
the black hole case. More precisely, since this solution can be studied by means of the extended conformal Einstein field equations ---see \cite{GasVal17a}. These equations are used to obtain a result concerning the evolution of the region of this spacetime which is bounded by the Cosmological horizon known as the \emph{Cosmological region}. In particular, in analogy to the de Sitter-like case, the
Cosmological region has an asymptotic region admitting a smooth conformal
extension with a spacelike conformal boundary and there exists a
conformal representation in which the induced $3$-metric on the
conformal boundary $\mathscr{I}$ is homogeneous. Thus, it is possible
to integrate the extended conformal field equations along single
conformal geodesics ---see \cite{Fri03c, GarGasVal18}.

\medskip
In this review article, the discussion of the construction of a conformal Gaussian gauge system leading to a hyperbolic reduction of the conformal Einstein field equations in the de Sitter-like case \cite{MinVal21} and the sub-extremal Schwarzschild-de Sitter case  \cite{MinVal23} is revisited and presented in a coherent and contiguous way.

\subsection{Notations and conventions}
The signature convention for Lorentzian spacetime metrics will be $
 (-,+,+,+)$. In this article, the abstract index notation is used. Accordingly, the lowercase Latin indices $\{_a ,_b , _c ,
 . . .\}$ will denote spacetime abstract tensor indices and $\{_\bma ,_\bmb , _\bmc ,
 . . .\}$ will be used as spacetime frame indices taking the values ${
   0, . . . , 3 }$.  In this way, given a basis
$\{\bme_{\bma}\}$ a generic tensor is denoted by $T_{ab}$ while its
components in the given basis are denoted by $T_{\bma \bmb}\equiv
T_{ab}\bme_{\bma}{}^{a}\bme_{\bmb}{}^{b}$. The Greek indices
${}_\mu,\, {}_\nu,\ldots$ denote spacetime coordinate indices while
the indices ${}_\alpha,\, {}_\beta,\ldots$ denote spatial coordinate indices.
An index-free notation is used where convenient. 

\medskip
The  conventions for the curvature tensors are fixed by the relation
\[
(\nabla_a \nabla_b -\nabla_b \nabla_a) v^c = R^c{}_{dab} v^d.
\]

\section{Tools of conformal geometry}
The purpose of this section is to provide a brief summary of the
technical tools of conformal geometry that will be used in the analysis of the evolution of the spacetimes under consideration.

\medskip
Let $(\tilde{\mathcal{M}},\tilde{\bmg})$ be a vacuum spacetime satisfying the Einstein field equations with positive
Cosmological constant
\begin{equation}
\tilde{R}_{ab} =\lambda \tilde{g}_{ab}
\label{EFE}
\end{equation}
and let $\bmg$ denote an unphysical Lorentzian metric conformally related
to $\tilde\bmg$ via the relation
\[
\bmg = \Xi^2 \tilde\bmg
\]
with $\Xi$ a suitable conformal factor. The Levi-Civita connections
of the metrics $\bmg$ and $\tilde\bmg$ are denoted by $\bmnabla$ and
$\tilde{\bmnabla}$, respectively. The set of points for which
$\Xi=0$ is called the \emph{conformal boundary}.

\subsection{Weyl connections}
A Weyl connection is a torsion-free connection $\hat{\bmnabla}$ such
that
\[
\hat{\nabla}_a g_{bc} =-2 f_a g_{bc}.
\]
It follows from the above that the connections $\nabla_a$ and
$\hat{\nabla}_a$ are related to each other by
\begin{equation}
\hat{\nabla}_av^b-\nabla_av^b = S_{ac}{}^{bd}f_dv^c, \qquad
S_{ac}{}^{bd}\equiv \delta_a{}^b\delta_c{}^d +
\delta_a{}^d\delta_c{}^b-g_{ac}g^{bd},
\label{WeylToUnphysical}
\end{equation}
where $f_a$ is a fixed smooth covector and $v^a$ is an arbitrary
vector.  Given that 
\[
\nabla_a v^b -\tilde{\nabla}_a v^b =  S_{ac}{}^{bd} (\Xi^{-1} \nabla_a\Xi)v^c,
\]
one has that 
\[
\hat{\nabla}_av^b -\tilde{\nabla}_av^b =  S_{ac}{}^{bd}\beta_d v^c,
\qquad \beta_d \equiv f_d +\Xi^{-1}\nabla_d\Xi.
\]
In the following, it will be convenient to define
\begin{equation}
d_a \equiv \Xi f_a + \nabla_a \Xi. 
\label{Definition:CovectorD}
\end{equation}

\subsection{The frame version of the extended conformal Einstein
  field equations}
\label{Section:FrameXCFE}
The \emph{extended conformal Einstein field equations} constitute a conformal
representation of the vacuum Einstein field equations written in terms
of \emph{Weyl connections} ---see \cite{Fri98a}. These equations are formally
regular at the conformal boundary. Moreover, a solution to the extended conformal
equations implies, in turn, a solution to the vacuum Einstein field equations
away from the conformal boundary.

\medskip
Let $\{ \bme_\bma
\}$, $\bma=\bmzero,\ldots,\bmthree$ denote a $\bmg$-orthogonal frame
with associated coframe $\{ \bmomega^\bma \}$. Thus, one has that
\[
\bmg(\bme_\bma,\bme_\bmb)=\eta_{\bma\bmb}, \qquad \langle
  \bmomega^\bma,\bme_\bmb\rangle =\delta_\bmb{}^\bma.
\]
The frame formulation of the \emph{extended
conformal Einstein field equations} is obtained by definining the following \emph{zero-quantities}:
\begin{subequations}
\begin{eqnarray} 
&& {\Sigma}{}_\bma{}^\bmc{}_\bmb \bme_\bmc\equiv [\bme_\bma, \bme_\bmb] - (\hat{\Gamma}{}_\bma{}^\bmc{}_\bmb- \hat{\Gamma}{}_\bmb{}^\bmc{}_\bma)e_\bmc, \label{ecfe1}\\
&&{\Xi}{}^\bmc{}_{\bmd\bma\bmb}\equiv {R}{}^\bmc{}_{\bmd\bma\bmb} - {\rho}{}^\bmc{}_{\bmd\bma\bmb}, \label{ecfe2} \\
&&{\Delta}{}_{\bmc\bmd\bmb} \equiv \hat\nabla_\bmc \hat{L}{}_{\bmd\bmb} - \hat\nabla_\bmd
   \hat{L}{}_{\bmc\bmb} - d_\bma d{}^\bma{}_{\bmb\bmc\bmd}, \label{ecfe3} \\
&&\Lambda{}_{\bmb\bmc\bmd} \equiv \hat\nabla_\bma
  d^\bma{}_{\bmb\bmc\bmd}-f_\bma d{}^\bma{}_{\bmb \bmc \bmd}, \label{ecfe4}
\end{eqnarray}
\end{subequations}
where the components of the \emph{geometric curvature} $\hat{R}{}^\bmc{}_{\bmd\bma\bmb}$ and the
\emph{algebraic curvature} $\hat{\rho}{}^\bmc{}_{\bmd\bma\bmb}$ are given, respectively, by
\begin{eqnarray*}
&& {R}{}^\bmc{}_{\bmd\bma\bmb} \equiv  \partial_\bma (\hat{\Gamma}{}_\bmb{}^\bmc{}_\bmd)- \partial_\bmb (\hat{\Gamma}{}_\bma{}^\bmc{}_\bmd) + \hat{\Gamma}{}_\bmf{}^\bmc{}_\bmd(\hat{\Gamma}{}_\bmb{}^\bmf{}_\bma - \hat{\Gamma}{}_\bma{}^\bmf{}_\bmb) + \hat{\Gamma}{}_\bmb{}^\bmf{}_\bmd \hat{\Gamma}{}_\bma{}^\bmc{}_\bmf - \hat{\Gamma}{}_\bma{}^\bmf{}_\bmd \hat{\Gamma}{}_\bmb{}^\bmc{}_\bmf,\\
&& {\rho}{}^\bmc{}_{\bmd\bma\bmb} \equiv \Xi \hat{d}{}^\bmc{}_{\bmd\bma\bmb} + 2 {S}{}_{\bmd
   [\bma}{}^{\bmc\bme}\hat{ L}{}_{\bmb]\bme}.
\end{eqnarray*}
In terms of the zero-quantities \eqref{ecfe1}-\eqref{ecfe4}, the
\emph{extended conformal Einstein field equations} are given by
the conditions
\begin{equation}
 {\Sigma}{}_\bma{}^\bmc{}_\bmb\bme_\bmc=0, \qquad {\Xi}{}^\bmc{}_{\bmd\bma\bmb}=0,
 \qquad  {\Delta}{}_{\bmc\bmd\bmb}=0, \qquad
 \Lambda{}_{\bmb\bmc\bmd}=0. \label{ecfe5}
\end{equation}
In the above equations the fields $\Xi$ and $d_\bma$ are regarded as \emph{conformal gauge
  fields} which are determined by gauge conditions. These conditions will be determined through
conformal geodesics ---see Subsection
\ref{Subsection:ConformalGeodesics} below. In order to account for
this it is convenient to define
\begin{subequations}
\begin{eqnarray}
&& \delta_\bma  \equiv d_\bma -\Xi f_\bma -\hat{\nabla}_\bma\Xi, \label{Supplementary1}\\
&& \gamma_{\bma\bmb} \equiv \hat{L}_{\bma\bmb}
   -\hat{\nabla}_\bma(\Xi^{-1} d_\bmb) -\frac{1}{2}\Xi^{-1}
   S_{\bma\bmb}{}^{\bmc\bmd}d_\bmc d_\bmd +
   \frac{1}{6}\lambda\Xi^{-2}\eta_{\bma\bmb}, \label{Supplementary2}\\
&& \varsigma_{\bma\bmb} \equiv \hat{L}_{[\bma\bmb]}
   -\hat{\nabla}_{[\bma} f_{\bmb]}. \label{Supplementary3}
\end{eqnarray}
\end{subequations}
The conditions
\begin{equation}
\delta_\bma =0, \qquad \gamma_{\bma\bmb}=0, \qquad
\varsigma_{\bma\bmb}=0,
\label{XCFESupplementary}
\end{equation}
will be called the \emph{supplementary conditions}. They play a role
in relating the Einstein field equations to the extended conformal
Einstein field equations and also in the propagation of the constraints.

\medskip
The correspondence between the Einstein field equations and the
extended conformal Einstein field equations is given by the following
---see Proposition 8.3 in \cite{CFEBook}:

\begin{proposition}
  \label{Lemma:XCFEtoEFE}
Let  
\[
(\bme_\bma, \, \hat{\Gamma}_\bma{}^\bmb{}_\bmc,
\hat{L}_{\bma\bmb},\,d^\bma{}_{\bmb\bmc\bmd})
\]
 denote a solution to
the extended conformal Einstein field equations \eqref{ecfe5} for some
choice of the conformal gauge fields $(\Xi,\,d_\bma)$ satisfying the
supplementary conditions \eqref{XCFESupplementary}. Furthermore,
suppose that
\[
 \Xi\neq 0, \qquad \det
(\eta^{\bma\bmb}\bme_\bma\otimes\bme_\bmb)\neq0
\]
 on an open subset
$\mathcal{U}$. Then the metric
\[
\tilde{\bmg}= \Xi^{-2} \eta_{\bma\bmb} \bmomega^\bma\otimes\bmomega^\bmb
\]
is a solution to the Einstein field equations \eqref{EFE} on
$\mathcal{U}$. 
\end{proposition}

\subsection{Conformal geodesics}
\label{Subsection:ConformalGeodesics}
The gauge used to analyse the evolution of the spacetimes under consideration is based on the properties of the \emph{conformal geodesics}. Conformal
geodesics allow the use of \emph{conformal Gaussian systems} in which
a certain canonical conformal factor gives an \emph{a priori} knowledge of the location
of the conformal boundary. This is in contrast with other conformal
gauges in which the conformal factor is an unknown. 

\subsubsection{Basic definitions}
A \textbf{\em conformal geodesic}\index{conformal geodesic!definition} on a spacetime
$(\tilde{\mathcal{M}},\tilde{\bmg})$ is a pair
  $(x(\tau),\bmbeta(\tau))$ consisting of a curve $x(\tau)$ and a covector $\bmbeta(\tau)$ along $x(\tau)$
  satisfying the equations
\begin{subequations}
\begin{eqnarray}
&& \tilde{\nabla}_{\dot{\bmx}} \dot{\bmx} = -2 \langle \bmbeta,
\dot{\bmx} \rangle
\dot{\bmx} + \tilde{\bmg}(\dot{\bmx},\dot{\bmx}) \bmbeta^\sharp, \label{ConformalGeodesicEquation1}\\
&& \tilde{\nabla}_{\dot{\bmx}} \bmbeta = \langle \bmbeta, \dot{\bmx}
\rangle \bmbeta - \displaystyle\frac{1}{2} \tilde{\bmg}^\sharp (\bmbeta,\bmbeta)
\dot{\bmx}^\flat + \tilde{\bmL}(\dot{\bmx},\cdot), \label{ConformalGeodesicEquation2}
\end{eqnarray}
\end{subequations}
where $\tilde{\bmL}$ denotes the \emph{Schouten tensor} of the
Levi-Civita connection $\tilde{\bmnabla}$.
A vector $\bmv$ is said to
be \emph{Weyl propagated} if along $x(\tau)$ it satisfies the equation
\begin{equation}
\tilde{\nabla}_{\dot{\bmx}} \bmv = -\langle \bmbeta,\bmv\rangle
\dot{\bmx} -\langle \bmbeta,\dot{\bmx}\rangle \bmv +
\tilde{\bmg}(\bmv,\dot{\bmx}) \bmbeta^\sharp.
\label{WeylPropagation}
\end{equation}

\smallskip
A congruence of conformal geodesics can be used to single out a metric
$\bmg\in[\tilde{\bmg}]$. This is due to the following property:
\begin{proposition}
  \label{ConformalFactor}
Let $(\tilde{\mathcal{M}},\tilde{\bmg})$ denote a vacuum spacetime with positive Cosmological constant. Suppose that $(x(\tau),\bmbeta(\tau))$ is a solution to the
conformal geodesic equations
\eqref{ConformalGeodesicEquation1}-\eqref{ConformalGeodesicEquation2}
and that $\{ \bme_\bma \}$ is a $\bmg$-orthonormal frame propagated
along the curve according to equation \eqref{WeylPropagation}. If
$\Theta$ satisfies
\begin{equation}
\bmg(\dot{\bmx},\dot{\bmx})=-1, \qquad \bmg=\Theta^2\tilde{\bmg},
\label{CG:NormalisationCondition}
\end{equation}
then one has
that
\begin{equation}
\Theta(\tau) = \Theta_\star + \dot{\Theta}_\star (\tau-\tau_\star)
+\frac{1}{2}\ddot{\Theta}_\star (\tau-\tau_\star)^2,
\label{CanonicalConformalFactorTheta}
\end{equation}
where the coefficients
\[
\Theta_\star\equiv \Theta(\tau_\star), \qquad 
\dot{\Theta}_\star\equiv \dot{\Theta}(\tau_\star)\qquad
\ddot{\Theta}_\star \equiv \ddot{\Theta}_\star(\tau_\star)
\]
are constant along the conformal geodesic and are subject to the
constraints
\[
\dot{\Theta}_\star = \langle \bmbeta_\star,\dot{\bmx}_\star\rangle
\Theta_\star, \qquad \Theta_\star \ddot{\Theta}_\star =
\frac{1}{2}\tilde{\bmg}^\sharp (\bmbeta_\star,\bmbeta_\star) + \frac{1}{6}\lambda.
\]
  \end{proposition}
A proof of this result can be found in \cite{CFEBook}.

\smallskip
Thus, if a spacetime can be covered by a non-intersecting congruence
of conformal geodesics, then the location of the conformal boundary is
known \emph{a priori} in terms of data at a fiduciary initial
hypersurface $\mathcal{S}_\star$.

\subsubsection{The $\tilde{g}$-adapted conformal geodesic equations}
As a consequence of the normalisation condition
\eqref{CG:NormalisationCondition}, the parameter $\tau$ is the
$\bmg$-proper time of the curve $x(\tau)$. In some computations it is
more convenient to consider a parametrisation in terms of a
$\tilde{\bmg}$-proper time $\tilde{\tau}$ of the curve $\tilde{x}(\tilde{\tau})$ as it allows to work
directly with 
the physical metric. To this end, consider the
parameter transformation $\tilde{\tau}=\tilde{\tau}(\tau)$ given by
\begin{equation}
\frac{\mbox{d}\tau}{\mbox{d}\tilde{\tau}} = \Theta,\qquad \mbox{so that}
\qquad \tilde{\tau} = \tilde{\tau}_\star + \int_{\tau_\star}^{\tau}
\frac{\mbox{ds}
  }{\Theta(\mbox{s})},
\label{CG:ChangeOfParameterFormula}
\end{equation}
with inverse $\tau=\tau(\tilde{\tau})$.
Now, consider 
\begin{equation}
\tilde{\bmx}'= \Theta
\dot{\bm x}, \qquad
\bmbeta=\tilde{\bmbeta} + \varpi \dot{\bmx}^\flat, \qquad \varpi \equiv 
\frac{\langle \bmbeta,
  \dot{\bmx}\rangle}{\tilde{\bmg}(\dot{\bmx},\dot{\bmx})},
\label{CG:OneFormSplit}
\end{equation}
in equations
\eqref{ConformalGeodesicEquation1}-\eqref{ConformalGeodesicEquation2}
so that one obtains the following $\tilde{\bm g}$-\textbf{\em adapted
  equations for the conformal geodesics}: 
\index{conformal geodesic!$\tilde{\bmg}$-adapted equations}
\begin{subequations}
\begin{eqnarray}
&& \tilde{\nabla}_{\tilde{\bmx}'} \tilde{\bmx}' = \tilde{\bmbeta}^\sharp, \label{PhysicalMetricAdaptedCC1}\\
&& \tilde{\nabla}_{\tilde{\bmx}'} \tilde{\bmbeta} = \tilde\beta^2 \tilde{\bmx}'{}^\flat + \tilde{\bmL}(\tilde{\bmx}',\cdot) -\tilde{\bmL}(\tilde{\bmx}',\tilde{\bmx}')\tilde{\bmx}^{\prime\flat}, \label{PhysicalMetricAdaptedCC2}
\end{eqnarray}
\end{subequations}
with $\tilde\beta^2\equiv \tilde{\bm
  g}^\sharp(\tilde{\bmbeta},\tilde{\bmbeta})$. For an vacuum spacetime with Cosmological constant one has that
\[
\tilde{\bmL}=\frac{1}{6}\lambda\tilde{\bmg}.
\]

\subsection{A Conformal Gaussian gauge system}
One considers a region $\mathcal{U}$ of the spacetime
$(\tilde{\mathcal{M}},\tilde{\bmg})$ which is covered by a non-intersecting
congruence of conformal geodesics
$({\bmx}(\tau),\bmbeta(\tau))$. The property of these curves stated by Proposition
   \ref{ConformalFactor} allows to single out a \emph{canonical
   representative} $\bmg$ of the conformal class $[\tilde{\bmg}]$ with
 an explicitly known conformal factor as given by the formula
 \eqref{CanonicalConformalFactorTheta}.

\smallskip
 Now, let $\{ \bme_\bma \}$ denote a $\bmg$-orthonormal frame which is
 Weyl propagated along the conformal geodesics. To every congruence of conformal geodesics
 one can associate a Weyl connection $\hat{\nabla}_a$ by setting $f_a=\beta_a$. It follows that
 for this connection one has
 \[
 \hat{\Gamma}_\bmzero{}^\bma{}_\bmb=0, \qquad f_\bmzero=0, \qquad \hat{L}_{\bmzero \bma}=0.
 \]
 This gauge choice is supplemented by choosing the parameter $\tau$
 of the conformal geodesics as the time coordinate so that
 \[
 \bme_\bmzero= \bmpartial_\tau. 
\]
Since the initial data for the
congruence of conformal geodesics is prescribed on a fiduciary spacelike
hypersurface $\mathcal{S}_\star$. On $\mathcal{S}_\star$ one can
choose some local coordinates $\underline{x}=(x^\alpha)$. These coordinates
can be extended off $\mathcal{S}_\star$ by requiring them to remain
constant along the conformal geodesic which intersects
$\mathcal{S}_\star$ at the point $p$ with
coordinates $\underline{x}$. The spacetime coordinates
$\overline{x}=(\tau,x^\alpha)$ obtained in this way are known as
\emph{conformal Gaussian coordinates}. 

\smallskip
The collection of conformal factor $\Theta$, Weyl propagated frame $\{ \bme_\bma \}$
and coordinates $(\tau,x^\alpha)$ is known as a \emph{conformal Gaussian gauge
  system}.

\subsection{The conformal constraint equations}
\label{Subsection:ConformalConstraints}
The \emph{conformal
  constraint Einstein equations}
are intrinsic equations implied by the conformal
Einstein field equations on a spacelike hypersurface. 

\medskip
Let $\mathcal{S}$ denote a spacelike hypersurface in an unphysical
spacetime $(\mathcal{M},\bmg)$ and let $\{ \bme_\bma \}$ denote a $\bmg$-orthonormal
frame adapted to $\mathcal{S}$. The conformal constraint equations in the vacuum case are given by
---see \cite{CFEBook}:
\begin{subequations}
\begin{eqnarray}
&& \label{co1}  D_\bmi D_\bmj \Omega =  \Sigma \chi_{\bmi\bmj} - \Omega L_{\bmi\bmj} + s h_{\bmi\bmj}, \\
&& \label{co2} D_\bmi \Sigma= {\chi_\bmi}^\bmk D_\bmk \Omega - \Omega L_\bmi, \\
&& \label{co3} D_\bmi s=  L_\bmi \Sigma - L_{\bmi\bmk} D^\bmk \Omega, \\
&&\label{co4} D_\bmi L_{\bmj\bmk} - D_\bmj L_{\bmi\bmk}=  \Sigma d_{\bmk\bmi\bmj} + D^\bml \Omega d_{\bml\bmk\bmi\bmj} - (\chi_{\bmi\bmk} L_\bmj - \chi_{\bmj\bmk} L_\bmi), \\
&&\label{co5} D_\bmi L_\bmj - D_\bmj L_\bmi=  D^\bml \Omega d_{\bml\bmi\bmj} +{\chi_\bmi}^\bmk L_{\bmj\bmk} - {\chi_\bmj}^\bmk L_{\bmi\bmk}, \\
&&\label{co6} D^\bmk d_{\bmk\bmi\bmj}= - ({\chi^\bmk}_\bmi d_{\bmj\bmk} - {\chi^\bmk}_\bmj d_{\bmi\bmk}), \\
&&\label{co7} D^\bmi d_{\bmi\bmj}=\chi^{\bmi\bmk} d_{\bmi\bmj\bmk}, \\
&&\label{co8} \lambda= 6 \Omega s + 3 \Sigma^2 - 3 D_\bmk \Omega D^\bmk \Omega, \\
&&\label{co9} D_\bmj \chi_{\bmk\bmi} - D_\bmk \chi_{\bmj\bmi} =\Omega d_{\bmi\bmj\bmk} + h_{\bmi\bmj} L_\bmk - h_{\bmi\bmk} L_\bmj, \\
&&\label{co10}l_{\bmi\bmj}= \Omega d_{\bmi\bmj} + L_{\bmi\bmj} - \chi (
 \chi_{\bmi\bmj} - \frac{1}{4} \chi h_{\bmi\bmj} ) + \chi_{\bmk\bmi}{\chi_\bmj}^\bmk -
\frac{1}{4}\chi_{\bmk\bml}\chi^{\bmk\bml} h_{\bmi\bmj},
\end{eqnarray}
\end{subequations}
with the understanding that 
\[
h_{\bmi\bmj}\equiv g_{\bmi\bmj}=\delta_{\bmi\bmj}
\]
 and where 
\[
L_\bmi\equiv L_{\bmzero\bmi}, \qquad d_{\bmi\bmj}\equiv
d_{\bmzero\bmi\bmzero\bmj}, \qquad d_{\bmi\bmj\bmk}\equiv d_{\bmi\bmzero\bmj\bmk}.
\]
Moreover, $\Omega$ denotes the restriction of the spacetime conformal
factor $\Xi$ to $\mathcal{S}$ and $\Sigma$ is the normal component of
the gradient of $\Xi$. The field $l_{\bmi\bmj}$ denotes the components of the
Schouten tensor of the induced metric $h_{ij}$ on $\mathcal{S}$. 
The fields $d_{\bmi\bmj}$ and $d_{\bmi\bmj\bmk}$ correspond,
respectively, to the electric and magnetic parts of the rescaled Weyl
tensor. The scalar $s$ denotes the \emph{Friedrich scalar} defined as
\[
s \equiv \frac{1}{4}\nabla_a\nabla^a \Xi + \frac{1}{24}R \Xi,
\]
with $R$ the Ricci scalar of the metric $\bmg$. Finally,
$L_{\bmi\bmj}$ denote the spatial components of the Schouten tensor of
$\bmg$.

\section{de Sitter-like spacetimes}
In this section, we 
study the evolution of de Sitter-like spacetimes which can be conformally embedded
into a portion of a cylinder whose sections have negative scalar
curvature as in \cite{MinVal21}. The conformal embedding is realised by means of a conformal
factor which depends on the affine parameter
of the conformal geodesics.

\subsection{Basic properties}
A de Sitter-like spacetime $(\tilde{\mathcal{M}},\mathring{\tilde{\bmg}})$
is a solution to the vacuum Einstein field equations with
positive Cosmological constant \eqref{EFE} given by  $\tilde{\mathcal{M}}=\mathbb{R}\times\mathcal{S}$ and
\begin{equation}	
\mathring{\tilde{\bmg}} = -\mathbf{d} t \otimes \mathbf{d} t + \sinh^2 t \; \mathring{\bmgamma},
\label{BackgroundSPhysicalMetric}
\end{equation}
where $\mathring{\bmgamma}$ is a positive definite Riemannian metric
over a compact manifold $\mathcal{S}$ with constant negative curvature.
\noindent
The Riemann curvature tensor $r^i{}_{jkl}[\mathring{\bmgamma}]$ of $\mathring{\bmgamma}$ is given by
\[
r_{ijkl}[\mathring{\gamma}] =
\mathring{\gamma}_{il}\mathring{\gamma}_{jk} - \mathring{\gamma}_{ik}\mathring{\gamma}_{jl}.
\]
 In particular, by setting $\lambda =3$, it follows from the above expressions that
\[
r[\mathring{\bmgamma}]=-6.
\]
Morever, since
\[
\tilde{R}=12,
\]
it follows that
\begin{equation}
\tilde{L}_{ab} = \frac{1}{2}\tilde{g}_{ab}.
\label{PhysicalSBackgroundSchouten}
\end{equation}
A spacetime of the form given by 
$(\tilde{\mathcal{M}},\mathring{\tilde{\bmg}})$ will be known as a
\emph{background solution}. 

\smallskip
{\em The value $\lambda=3$ for the Cosmological constant is
  conventional and set for convenience. This analysis
  can be carried out for any other positive value of $\lambda$. }

\subsection{Metric geodesics as conformal geodesics}
The analysis of the metric geodesics $x(s)$ on $(\tilde{\mathcal{M}},\mathring{\tilde{\bmg}})$ with
$\dot{\bmx} \equiv \frac{\partial x}{\partial s} = \alpha \bmpartial_t$, where $\alpha$ is a proportionality function, by means of the geodesic equation
\[
\tilde{\nabla}_{\dot{\bmx}} \dot{\bmx}=0
\]
and the metric \eqref{BackgroundSPhysicalMetric}
shows that $\alpha$ is constant along the integral curves of
$\bmpartial_t$. Hence, without loss of generality one can set
$\alpha=1$ so that the curves
\[
x(t) = (t, \underline{x}_\star), \qquad \underline{x}_\star\in \mathcal{S},
\]
are non-intersecting timelike $\tilde{\bmg}$-geodesics over
$\tilde{\mathcal{M}}$. These curves can be recasted as conformal
  geodesics by means of a reparametrisation 
$\tau \mapsto t(\tau)$ and a 1-form $\tilde{\bmbeta}$ given by the Ansatz
\[
\tilde\bmbeta = \alpha(\tau) {\bmx}^{\prime\flat} = \alpha(t) \mathbf{d}t.
\]
The resulting pair $(x(\tau),\tilde{\bmbeta}(\tau))$ with
\[
x(\tau) = (2\,\mbox{arctanh}\,\tau, \underline{x}_\star), \qquad
\tilde{\bmbeta}(\tau) = -\frac{2\tau}{1-\tau^2}\mathbf{d}\tau \qquad {\rm and} \hspace{0.5cm} \tau\in(-1,1)
\]
describes a congruence of non-intersecting timelike conformal geodesics on
the background spacetime
$(\tilde{\mathcal{M}},\mathring{\tilde{\bmg}})$. 

\subsection{The conformal factor associated to the congruence of
  conformal geodesics}
The parameter $\tau$ introduced in the previous section is used as a new time coordinate in the metric
\eqref{BackgroundPhysicalMetric} so that
\begin{equation}
\mathring{\tilde{\bmg}}  = \frac{4}{(1-\tau^2)^2}\bigg(
-\mathbf{d}\tau\otimes\mathbf{d}\tau +\tau^2 \mathring{\bmgamma}  \bigg).
\label{ReparametrisedMetric}
\end{equation}
This metric is singular at $\tau=\pm 1$. 
This line element suggest the introduction of a new
unphysical metric $\mathring{\bmg}$ via the relation
\[
 \mathring{\bmg} = \Theta^2 \mathring{\tilde{\bmg}}, \qquad {\rm with} \hspace{0.5cm} \Theta \equiv \frac{1}{2}{(1- \tau^2)},
\]
so that 
\begin{equation}
\mathring{\bmg}= -\mathbf{d} \tau \otimes \mathbf{d} \tau + \tau^2
\mathring{\bm{\gamma}} 
\label{BackgroundMetric}
\end{equation}
is well defined for $ \tau \in [ {\tau}_\star, \infty)$ with
$\tau_\star>0$. The \emph{spatial metric}
$\mathring\bmh$ is conformally related to $\mathring\bmgamma$ via
\[
\mathring\bmh \equiv \tau^2\mathring\bmgamma, 
\]
with associated Levi-Civita connection to be denoted by $\mathring
D$, whereas $\mathring{\mathfrak{D}}$ is the Levi-Civita connection
 of the metric $\mathring\bmgamma$. The integral curves of the
  vector field $\bmpartial_\tau$ are geodesics of the metric
  $\mathring\bmg$ given by equation \eqref{BackgroundMetric}.
Moreover, since $\tilde\bmbeta$ is a closed 1-form the
Weyl connection is, in fact, a Levi-Civita connection which coincides
with $\bmnabla$.

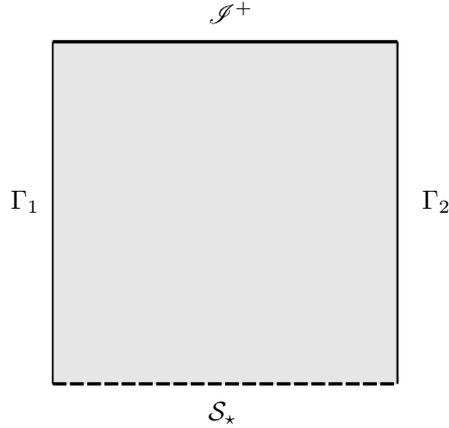
\begin{figure}[t]
\centering
\includegraphics[width=50mm]{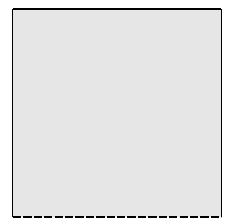}
\put(-76,140){$\mathscr{I}^+$}
\put(-76,-10){$\mathcal{S}_\star$}
\put(-150,70){$\Gamma_1$}
\put(4,70){$\Gamma_2$}
\caption{Penrose diagram of the background solution. The conformal
  representation discussed in the main text has compact sections of
  negative scalar curvature. The vertical lines $\Gamma_1$ and
  $\Gamma_2$ correspond to axes of symmetry. The solution has a singularity in the
  past and a spacelike future conformal boundary. Hence, in our discussion we only consider future evolution of
  the initial hypersurface $\mathcal{S}_\star$.}
\label{Figure:PenroseDiagram}
\end{figure}

A Penrose diagram of conformal
 representation of the background solution described by the metric
 \eqref{BackgroundMetric} is given in Figure \ref{Figure:PenroseDiagram}.

\subsection{The background spacetime as a solution to the conformal
  Einstein field equations}
The \emph{unphysical
  spacetime} $(\mathcal{M},\mathring{\bmg})$ is recasted as a solution to the conformal
Einstein field equations. This construction is done using
an adapted frame formalism. 

\subsubsection{The frame}
Let $\{\mathring{\bmc}_\bmi\}$, $\bmi=1,\, 2,\,3$, denote a
$\mathring{\bmgamma}$-orthonormal frame over $\mathcal{S}$ with associated cobasis $\{ \mathring{\bmalpha}^\bmi \}$. Accordingly, one has that 
\[
\mathring{\bm{\gamma}}(\mathring\bmc_i, \mathring\bmc_j)= \delta_{\bmi \bmj}, \qquad \langle \mathring\bmalpha^\bmj, \mathring\bmc_\bmi \rangle= \delta_\bmi{}^\bmj,
\]
so that
\[
\mathring{\bmgamma} = \delta_{\bmi\bmj} \mathring\bmalpha^\bmi \otimes \mathring\bmalpha^\bmj.
\] 
The above frame is used to introduce a $\mathring{\bmg}$-orthonormal
frame $\{ \mathring{\bme}_\bma \}$ with associated cobasis $\{
\mathring{\bmomega}^\bmb\}$ so that $\langle \mathring{\bmomega}^\bmb,
\mathring{\bme}_\bma\rangle =\delta_\bma{}^\bmb$. This is done by
setting
\begin{eqnarray*}
\mathring{\bme}_0 \equiv\bmpartial_\tau, &&
                                       \mathring{\bme}_\bmi\equiv \frac{1}{\tau}\mathring{\bmc}_\bmi,\\
\mathring{\bmomega}^0 \equiv \mathbf{d}\tau, &&
                                                \mathring{\bmomega}^\bmi
                                                = \tau \mathring{\bmalpha}^\bmi, 
\end{eqnarray*}
so that 
\[
\mathring{\bmg} = \eta_{\bma\bmb} \mathring{\bmomega}^\bma \otimes \mathring{\bmomega}^\bmb.
\]

\subsubsection{The connection coefficients}
The connection coefficients $\mathring\gamma_\bmi{}^\bmk{}_\bmj$ of
the Levi-Civita connection $\mathring{D}$ with respect to the frame
$\{\mathring\bmc_\bmi \}$ are defined through the relations
\[
\mathring D_\bmi \mathring\bmc_\bmj =\mathring\gamma_\bmi{}^\bmk{}_\bmj
\mathring\bmc_\bmk, \qquad \gamma_\bmi{}^\bmk{}_\bmj \equiv \langle
\mathring\bmalpha^\bmk, \mathring D_\bmi \mathring\bmc_\bmj \rangle.
\]
Similarly, for the connection coefficients
$\mathring\Gamma_\bmi{}^\bmk{}_\bmj$ of the Levi-Civita connection
$\mathring\bmnabla$ with respect to the frame $\{ \mathring\bme_\bma \}$
one has that 
\[
\mathring \nabla_\bma \mathring\bme_\bmb =
\mathring\Gamma_\bma{}^\bmc{}_\bmb \mathring\bme_\bmc, \qquad
\mathring\Gamma_\bma{}^\bmc{}_\bmb \equiv \langle \mathring\bmomega^\bmc, \mathring \nabla_\bma \mathring\bme_\bmb\rangle.
\]
Using these relations, it follows that the only non-vanishing connection coefficients are
\begin{equation} \label{ConnectionCoefficients}
\mathring\Gamma_\bmi{}^\bmk{}_\bmj =\frac{1}{\tau}\mathring\gamma_\bmi{}^\bmk{}_\bmj, \qquad \mathring\Gamma_\bmi{}^\bmj{}_\bmzero =\mathring\chi_\bmi{}^\bmj, \qquad \mathring{\Gamma}_\bmi{}^\bmzero{}_\bmj  =\frac{1}{\tau} \delta_{\bmi\bmj}, \qquad \mathring\Gamma_\bmzero{}^\bmj{}_\bmi = -\frac{1}{\tau}\delta_\bmi{}^\bmj,
\end{equation}
where $\mathring{\chi}_\bmi{}^\bmj$ denote the components of the \emph{Weingarten
  tensor}. 

Thus, all the connection coefficients are smooth over $[\tau_\star,\infty)\times\mathcal{S}$.

\subsubsection{Conformal fields}
The components of the conformal
fields appearing in the extended conformal Einstein field
equations are obtained by solving the conformal Einstein
constraints discussed in Section \ref{Subsection:ConformalConstraints}.

\medskip
This is done by means of an adapted frame with
$\bme_0=\bmpartial_\tau$ and by making the identification $\Omega\mapsto
\Theta$ in equations \eqref{co1}-\eqref{co10}. The analysis of these equations gives
\begin{subequations}
\begin{eqnarray}
&&\mathring{D}_i \Omega =0, \qquad \mathring{\Sigma} \equiv \bmn (\Theta) = \tau, \qquad \mathring{s}=1,  \\
&&\mathring{L}_i=0, \qquad  \mathring{L}_{ij}=0, \qquad \mathring{d}{}^*{}_{ij}=0, \qquad \mathring{d}{}_{ij}=0.
\end{eqnarray}
\end{subequations}
Thus, all the fields are regular up to the conformal boundary and the metric \eqref{BackgroundMetric} is
conformally flat.

\subsection{Evolution equations}
\label{Section:EvolutionEqns}
In this section we discuss the evolution system associated to the
extended conformal Einstein equations \eqref{ecfe5} written in terms
of a conformal Gaussian system. In addition, we also
discuss the subsidiary evolution system satisfied by the
zero-quantities associated to the field equations, \eqref{ecfe1}-\eqref{ecfe4},
and the supplementary zero-quantities \eqref{Supplementary1}-\eqref{Supplementary3}. 

\subsubsection{The conformal Gaussian gauge}
To obtain suitable evolution equations for the conformal
fields it is used a \emph{conformal Gaussian gauge}. More
precisely, it is assumed that a region
$\mathcal{U}\subset\mathcal{M}$ is covered by a congruence
of non-intersecting conformal geodesics. Then, by choosing
\[
\Theta_\star =\frac{1}{2}, \qquad \dot{\Theta}_\star =0, \qquad
\ddot{\Theta}_\star =-\frac{1}{2},
\]
for $\tau=\tau_\star$, $\tau_\star \in(0,1)$, Proposition \ref{ConformalFactor} gives the
conformal factor
\begin{equation}
\Theta(\tau) =\frac{1}{2}\big( 1-(\tau-\tau_\star)^2\big)
\label{UniversalCF}
\end{equation}
along the curves of the congruences. The choice of initial data for the
conformal factor is associated to a congruence that leaves
orthogonally a fiduciary initial hypersurface $\mathcal{S}_\star$ with
$\tau=\tau_\star$. Since the conformal factor $\Theta$ given by equation
  \eqref{UniversalCF} does not depend on the initial data for the
  evolution equations it can be regarded as valid
  not only for the background solution but also for its perturbations. 

\smallskip
Along the congruence of conformal geodesics one considers a
$\bmg$-orthogonal frame $\{\bme_\bmzero\}$ which is Weyl-propagated
and such that $\bmtau=\bme_\bmzero$. The Weyl connection $\hat{\nabla}_a$ associated to
the congruence then satisfies
\[
\hat{\nabla}_\bmtau \bme_\bma=0, \qquad \hat{\bmL}(\tau,\cdot)=0.
\]
 By choosing the parameter, $\tau$, of the conformal geodesics as time
coordinate one gets the additional gauge condition 
\[
\bme_\bmzero = \bmpartial_\tau, \qquad e_\bmzero{}^\mu =\delta_0{}^\mu.
\]
On $\mathcal{S}_\star$ we choose some local coordinates
$\underline{x}=(x^\alpha)$. These coordinates can be extended
 off the initial hypersurface so that the
coordinates $(\tau,\underline{x})$ thus obtained are
\emph{conformal Gaussian coordinates}.

\subsubsection{Structural properties of the evolution and subsidiary equations}
\label{Subsection:SubsidiarySystem}
In the conformal Gaussian gauge, the various fields associated to the extended
vacuum conformal Einstein field equations satisfy the evolution equations
\begin{subequations}
\begin{eqnarray}
&& \partial_\tau e_\bmb{}^\nu = - \hat{\Gamma}{}_\bmb{}^\bmc{}_\bmzero e_\bmc{}^\nu, \label{XCFEEvolution1}\\
&&\partial_\tau \hat{L}{}_{\bmd\bmb}=\hat{\Gamma}{}_\bmzero{}^\bmc{}_\bmd \hat{L}{}_{\bmc\bmb}+ \hat{\Gamma}{}_\bmzero{}^\bmc{}_\bmb \hat{L}{}_{\bmd\bmc} + d_\bma \hat{d}{}^\bma{}_{\bmb\bmzero\bmd}, \label{XCFEEvolution2}\\
&&  \partial_\tau f_\bmi = - f_\bmj \hat{\Gamma}{}_\bmi{}^\bmj{}_\bmzero + \hat{L}{}_{\bmi\bmzero}, \label{XCFEEvolution3}\\
&& \partial_\tau (\hat{\Gamma}{}_\bmb{}^\bmc{}_\bmd) = -\hat{\Gamma}{}_\bmf{}^\bmc{}_\bmd \hat{\Gamma}{}_\bmb{}^\bmf{}_\bmzero-\Xi \hat{d}{}^\bmc{}_{\bmd\bmzero\bmb}-2{\delta_\bmd}^\bmc \hat{L}_{\bmb\bmzero}-2{\delta_\bmzero}^c \hat{L}_{\bmb\bmd}+2 g_{\bmd\bmzero}g^{\bmc\bme}\hat{L}_{\bmb\bme},\label{XCFEEvolution4}\\
&& \partial_\tau d_{\bmb\bmd}+\epsilon^{\bme\bmf}{}_{(\bmd}D_\bmf d^*{}_{\bmb)\bme}= 2a_\bmf\epsilon^{\bme\bmf}_{(\bmd} d^*{}_{\bmb)\bme}- \chi d_{\bmb\bmd} +2 \chi^\bmf{}_{(\bmb} d_{\bmd)\bmf}, \label{XCFEEvolution5}\\
&&  \partial_\tau d^*{}_{\bmb\bmd} -\epsilon^\bme{}_{\bmf(\bmd}D^\bmf
   d_{\bmb)\bme} = 2 a^\bmf {\epsilon_{\bmf(\bmd}}^\bme d_{\bmb)\bme} - \chi
   d^*{}_{\bmb\bmd} + 2 \chi^\bmf_{(\bmb} d^*{}_{\bmd)\bmf}. \label{XCFEEvolution6}
\end{eqnarray}
\end{subequations}
Letting $\bme$, $\bmGamma$, $\hat{\bmL}$ and $\bmphi$ denote,
respectively, the independent components of the coefficients of the
frame, the connection coefficients, the Schouten tensor of the Weyl
connection and the rescaled Weyl tensor and setting, for convenience,
${\mathbf{u}}\equiv ( {\bmupsilon}, {\bmphi})$ with
${\bmupsilon}\equiv  (\bme, {\bmGamma}, {\bmL})$ and ${\bmphi}\equiv  ({\bmd}, {\bmd}^*)$ one has the following:
\begin{lemma}
  \label{Lemma:EvolutionEqns}
The extended conformal Einstein field equations \eqref{ecfe5}
expressed in in terms of a conformal Gaussian gauge imply that the evolution equations \eqref{XCFEEvolution1}-\eqref{XCFEEvolution6} can be written as a symmetric
hyperbolic system for the components $({\bmupsilon},{\bmphi})$ of
the form
\begin{subequations}
\begin{align}
&\partial {\bmupsilon} = \mathbf{K} {\bmupsilon} +
   \mathbf{Q}({\bmGamma}){\bmupsilon} + \mathbf{L}(\bar{x}) \bmphi, \label{he1}\\
  & \big( \mathbf{I} + \mathbf{A}^0(\bme)  \big)\partial_\tau \bmphi
     + \mathbf{A}^\alpha(\bme) \partial_\alpha \bmphi
     =\mathbf{B}({\bmGamma})\bmphi, \label{he2}
     \end{align}
     \end{subequations}
where $\mathbf{I}$ is the unit matrix, $\mathbf{K}$ is a constant
matrix $\mathbf{Q}({\bmGamma})$ is a smooth matrix-valued
function, $\mathbf{L}(\bar{x})$ is a smooth matrix-valued
function of the coordinates, $\mathbf{A}^\mu(\bme)$ are Hermitian
matrices depending smoothly on the frame coefficients and
$\mathbf{B}({\bmGamma})$ is a smooth matrix-valued function of the
connection coefficients.
\end{lemma}
Regarding the subsidiary evolution system, it follows from
the system
\begin{subequations}
\begin{eqnarray}
&& \hat{\nabla}_{\bmzero} \hat{\Sigma}{}_{\bmb}{}^\bmd{}_{\bmc}=-
\frac{1}{3}\hat{\Gamma}{}_\bmc{}^\bme{}_\bmzero \hat{\Sigma}{}_\bme{}^\bmd{}_\bmb -
\frac{1}{3}\hat{\Gamma}{}_\bmc{}^\bme{}_\bmzero\hat{\Sigma}{}_\bme{}^\bmd{}_\bmb -
\hat{\Xi}{}^\bmd{}_{\bmzero\bmb\bmc}, \label{SubsidaryEqnFirst}\\
&& \hat{\nabla}_\bmzero \hat{\Xi}{}^\bmd{}_{\bme\bmb\bmc}=\hat{\Gamma}{}_\bmb{}^\bmf{}_\bmzero \hat{\Xi}{}^\bmd{}_{\bme\bmc\bmf} + \hat{\Gamma}{}_\bmc{}^\bmf{}_\bmzero \hat{\Xi}{}^\bmd{}_{\bme\bmf\bmb}- \hat{\Sigma}{}_\bmb{}^\bmf{}_\bmc \hat{R}{}^\bmd{}_{\bme\bmzero\bmf} - \frac{1}{2} \Theta \epsilon{}^\bmf{}_{\bmzero\bmb\bmc} \epsilon{}_\bme{}^{\bmd\bmg\bmh}\Lambda{}_{\bmf\bmg\bmh} \\
&& \hspace{2cm} + \epsilon{}^\bmf{}_{\bmzero\bmb\bmc} \delta^\bmg d{}^{*\bmd}{}_{\bme\bmf\bmg} + 3S{}_{\bme\bmzero}{}^{\bmd\bmg}
\hat{\Delta}{}_{\bmc\bmb\bmg}, \\
&& \hat{\nabla}_\bmzero \hat{\Delta}_{\bmb\bmc\bmd}= \hat{\Gamma}{}_\bmb{}^\bme{}_\bmzero
\hat{\Delta}{}_{\bmc\bme\bmd}+ \hat{\Gamma}{}_\bmc{}^\bme{}_\bmzero \hat{\Delta}{}_{\bme\bmb\bmd}-
\hat{\Xi}{}^\bme{}_{\bmzero\bmb\bmc} \hat{L}{}_{\bme\bmd} + \delta_\bmb d_\bme d{}^\bme{}_{\bmd\bmc\bmzero} +
\delta_\bmc d_\bme d{}^\bme{}_{\bmd\bmzero\bmb} \\
&&  \hspace{2cm}+ \Theta \gamma{}_{\bmb\bme} d{}^\bme{}_{\bmd\bmc\bmzero} +
\Theta \gamma{}_{\bmc\bme} d{}^\bme{}_{\bmd\bmzero\bmb} - \frac{1}{2}\epsilon{}_{\bmzero\bmb\bmc}{}^\bmf
\epsilon{}_\bmd{}^{\bme\bmg\bmh} \Lambda{}_{\bmf\bmg\bmh} \beta_\bme, \\
&& \hat{\nabla}{}_\bmzero \hat{\Omega}{}_{\bmb\bmc} = \hat{\Xi}{}^\bme{}_{[\bmb}{}^{\bma\bmf}
d{}_{\bmc]\bme\bmf\bma} - \hat{\Xi}{}^\bme{}_\bmf{}^{\bma\bmf} d{}_{\bme\bma\bmb\bmc} + \frac{1}{2}
\hat{\Sigma}{}_\bma{}^\bme{}_\bmf \nabla_\bme d{}^{\bmf\bma}{}_{\bmb\bmc} +
\varsigma{}^{\bmf\bma}d{}_{\bmf\bma\bmb\bmc}-\chi \Omega{}_{\bmb\bmc},\\
&&\hat{\nabla}_\bmzero \delta_\bmi= \gamma_{\bmi\bmzero}- \hat{\Gamma}{}_\bmi{}^\bme{}_\bmzero \delta_\bme; \\
&& \hat{\nabla}_\bmzero \gamma_{\bmi\bmc}= - \gamma_{\bmj\bmc} \hat{\Gamma}{}_\bmi{}^\bmj{}_\bmzero
   - \beta_\bmzero \gamma_{\bmi\bmc} - \beta_\bmc \gamma_{\bmi\bmzero} + \eta_{\bmzero\bmc}(\beta^\bme
   \gamma_{\bmi\bme} - 2 \lambda \Theta^{-2} \delta_\bmi), \\
&& \hat{\nabla}_\bmzero \varsigma_{\bmj\bmk} =
   \hat{\Gamma}{}_\bmj{}^\bme{}_\bmzero \varsigma{}_{\bmk\bme} +
   \hat{\Gamma}{}_\bmk{}^\bme{}_\bmzero \varsigma{}_{\bme\bmj} + \frac{1}{2}
   \hat{\Delta}_{\bmj\bmk\bmzero} + \frac{1}{2}
   \hat{\Xi}{}^\bme{}_{\bmzero\bmj\bmk} f_\bme + \frac{1}{2}
   \hat{\Sigma}{}_\bmj{}^\bme{}_\bmk
   \hat{\Gamma}{}_\bme{}^\bmf{}_\bmzero f_\bmf, \label{SubsidaryEqnLast}
\end{eqnarray}
\end{subequations}
that the zero-quantities
$\hat{\Sigma}_\bma{}^\bmc{}_\bmb$, $\hat{\Xi}^\bma{}_{\bmb\bmc\bmd}$,
$\hat{\Delta}_{\bma\bmb\bmc}$, $\hat{\Lambda}_{\bma\bmb\bmc}$,
$\delta_{\bma\bmb}$, $\gamma_{\bma\bmb}$ and $\varsigma_{\bma\bmb}$
satisfy, if the conformal evolution equations
\eqref{XCFEEvolution1}-\eqref{XCFEEvolution5} hold, a symmetric
hyperbolic system which is homogeneous in the zero-quantities. 
More precisely, upon defining
$\hat{\mathbf{X}}\equiv(\hat{\Sigma}{}_\bma{}^\bmc{}_\bmb,\hat{\Xi}{}^\bmc{}_{\bmd\bma\bmb},
\hat{\Delta}{}_{\bma\bmb\bmc}, \hat{\Lambda}{}_{\bma\bmb\bmc},
\delta_{\bma}, {\gamma}{}_{\bma\bmb}, {\varsigma}{}_{\bma\bmb})$,
these equations can be recasted as a symmetric hyperbolic system of
the form
\begin{equation}
\label{SubEv}
\partial_\tau \hat{\mathbf{X}}= \mathbf{H}(\hat{\mathbf{X}}),
\end{equation}
where $\mathbf{H}(\bmzero)=\bmzero$. The particular situation in which all the
zero-quantities vanish identically gives rise to the subsidiary
evolution system.

\subsection{A perturbative argument} 
In the following, we look for solutions to the system
\eqref{he1}-\eqref{he2} of the form 
\[
\hat{\mathbf{u}} = \mathring{\mathbf{u}} + \breve{\mathbf{u}} 
\]
where $\mathring{\mathbf{u}}$ is the solution to the conformal
evolution equations \eqref{XCFEEvolution1}-\eqref{XCFEEvolution6}
implied by a background solution, while $\breve{\bmu}$ denotes a small
perturbation. Accordingly, one can set
\begin{subequations}
\begin{eqnarray} 
&& \label{per1} \hat{\bmupsilon} = \mathring{\bmupsilon} + \breve{\bmupsilon}, \hspace{1cm} \hat{\bmphi} = \breve{\bmphi}, \\
&&  \label{per2} \hat{\bme} = \mathring{\bme} + \breve{\bme}, \hspace{1cm}  \hat{\bmGamma} = \mathring{\bmGamma} + \check{\bmGamma}.
\end{eqnarray}
\end{subequations}  
Now, on the initial surface $\mathcal{S}_\star$
described by the condition $\tau=\tau_\star$ one has that
$\mathring{\mathbf{u}}_\star =( \mathring{\bmupsilon}_\star,
\mathring{\bmphi}_\star) = ( \mathring{\bmupsilon}_\star, 0)$ being the exact de Sitter-like solution. As the
conformal factor $\Theta$ and the covector $\bmd$ are
universal, it follows that
\[
 \partial_\tau \mathring{\bmupsilon} = \mathbf{K} \mathring{\bmupsilon} + \mathbf{Q}( \mathring{\bmupsilon}, \mathring{\bmupsilon}). 
\]
Substituting \eqref{per1} and \eqref{per2} into equations \eqref{he1}
and \eqref{he2} and upon defining the following matrices
\[
\bar{\mathbf{A}}^0(\tau, \underline{x}, \breve{\mathbf{u}}) \equiv \begin{pmatrix*}
      {\rm I} &  0 \\
      0 & {\rm I} + \mathbf{A}^0(\mathring{\bme}+ \breve{\bme})
     \end{pmatrix*},
\qquad 
\bar{\mathbf{A}}^\alpha(\tau, \underline{x}, \breve{\mathbf{u}}) \equiv \begin{pmatrix*}
  0 &   0 \\
      0 &   \mathbf{A}^\alpha(\mathring{\bme}+ \breve{\bme})
     \end{pmatrix*}
\]
and
\[
\bar{\mathbf{B}}(\tau, \underline{x}, \breve{\mathbf{u}})\equiv \breve{\mathbf{u}}\bar{\mathbf{Q}}\breve{\mathbf{u}}+ \bar{\mathbf{L}}(\bar{x})\breve{\mathbf{u}}+ \bar{\mathbf{K}}\breve{\mathbf{u}},
\]
where 
\[
\breve{\mathbf{u}}\bar{\mathbf{Q}}\breve{\mathbf{u}} \equiv \begin{pmatrix*}
    \breve{\bmupsilon}\mathbf{Q}\breve{\bmupsilon} &     0 \\
     0 &
     \mathbf{B}(\breve{\bmGamma})\breve{\bmphi}
     \end{pmatrix*},
\qquad 
\bar{\mathbf{L}}(\bar{x})\breve{\mathbf{u}} \equiv \begin{pmatrix*}
  \mathring{\bmupsilon}\mathbf{Q}\breve{\bmupsilon}+ \mathbf{Q}(\breve{\bmGamma})\mathring{\bmupsilon}&
    \mathbf{L}(\bar{x})\breve{\bmphi} \\
      0 &  0
     \end{pmatrix*},
\qquad 
\bar{\mathbf{K}} \breve{\mathbf{u}}\equiv \begin{pmatrix*}
 \mathbf{K} \breve{\bmupsilon} &  0  \\
      0 & \mathbf{B}(\mathring{\Gamma})\breve{\bmphi}
     \end{pmatrix*},
\]
it is possible to write the evolution equations for $\breve{\mathbf{u}}=(\breve{\bmupsilon}, \breve{\bmphi})$ as
\begin{equation}
 \bar{\mathbf{A}}^0(\tau, \underline{x}, \breve{\mathbf{u}})\partial_\tau \breve{\mathbf{u}}+ \bar{\mathbf{A}}^\alpha(\tau, \underline{x}, \breve{\mathbf{u}})\partial_\alpha \breve{\mathbf{u}}=\bar{\mathbf{B}}(\tau, \underline{x}, \breve{\mathbf{u}}). \label{he5}
\end{equation}
Since this is a symmetric hyperbolic system, existence and stability results are obtained by using known results for
symmetric hyperbolic systems with compact spatial sections ---see
e.g. \cite{CFEBook}, Section 12.3 which, in turn, follow from Kato's
theory for symmetric hyperbolic systems over $\mathbb{R}^n$
\cite{Kat75c}. The existence and Cauchy stability of the solution to the initial
value problem for the original conformal evolution problem
\begin{eqnarray*} 
&& \mathbf{A}^0(\tau, \underline{x},
\hat{\mathbf{u}})\partial_\tau \hat{\mathbf{u}}+ \mathbf{A}^\alpha(\tau,
\underline{x}, \hat{\mathbf{u}})\partial_\alpha \hat{\mathbf{u}}=\mathbf{B}(\tau,
\underline{x}, \hat{\mathbf{u}}), \\ 
&& \hat{\mathbf{u}}|_\star=
\mathring{\mathbf{u}}_\star + \breve{\mathbf{u}}_\star \in H^m (\mathcal{S}_\star,
\mathbb{R}^N) \quad {\rm for} \quad m \geq 4
\end{eqnarray*} 
follows from the fact that $\hat{\mathbf{u}}$ satisfies the
same properties as $\breve{\mathbf{u}}$ and then it exists in
the same solution manifold and with the same regularity properties,
existence and uniqueness.

\subsection{A solution to the Einstein field equations}
\label{Section:PropagationofConstraints}
In this section, we discuss the connection
between the solution to the conformal evolution systems and the actual
solution to the Einstein field equations. 

\smallskip
From the discussion in 
Section \ref{Subsection:SubsidiarySystem} it follows that the independent components of the zero-quantities
$\hat{\mathbf{X}}\equiv(\hat{\Sigma}{}_\bma{}^\bmc{}_\bmb,\hat{\Xi}{}^\bmc{}_{\bmd\bma\bmb},
\hat{\Delta}{}_{\bma\bmb\bmc}, \hat{\Lambda}{}_{\bma\bmb\bmc},
\delta_{\bma}, {\gamma}{}_{\bma\bmb}, {\varsigma}{}_{\bma\bmb})$
satisfy the symmetric hyperbolic system \eqref{SubEv}. 
Then, a solution to the initial value problem
\begin{eqnarray*}
&& \partial_\tau \hat{\mathbf{X}}= \mathbf{H}(\hat{\mathbf{X}}), \\
&& \hat{\mathbf{X}}_\star =0
\end{eqnarray*}
is given by $\hat{\mathbf{X}}=0$. Moreover, from
Kato's theorem \cite{Kat75c} follows that this is the unique solution. Thus, the
zero-quantities must vanish on $\big{[}\tau_\star, 1\big{)}\times
\mathcal{S}_\star$. This result is summarised by the following 
\begin{proposition}[\textbf{\em propagation of the constraints}]
\label{Prop:PropagationofConstraints}
 Let $\hat{\mathbf{u}}_\star = \mathring{\mathbf{u}}_\star + \breve{\mathbf{u}}_\star$ denote
 initial data for the conformal evolution equations on a $3$-manifold
 $\mathcal{S}_\star$ such that 
\[
\hat{\Sigma}{}_\bma{}^\bmc {}_\bmb |_{\mathcal{S}_\star}=0, \quad
\hat{\Xi}{}^c {}_{\bmd\bma\bmb}  |_{\mathcal{S}_\star}=0, \quad
\hat{\Delta}{}_{\bma\bmb\bmc}  |_{\mathcal{S}_\star}=0, \quad  \hat{
  \Lambda}_{\bma\bmb\bmc} |_{\mathcal{S}_\star}=0, 
\]
and
\[
 \delta{}_\bma |_{\mathcal{S}_\star}=0, \quad \gamma{}_{\bma\bmb}
 |_{\mathcal{S}_\star}=0, \quad \varsigma{}_{\bma\bmb} |_{\mathcal{S}_\star}=0, 
\]
then the solution $\breve{\mathbf{u}}$ to the conformal evolution equations
implies a $C^{m-2}$ solution $\hat{\mathbf{u}}= \mathring{\mathbf{u}} +
\breve{\mathbf{u}}$ to the extended conformal field equations on $\big{[}
\tau_\star, 1 \big{)} \times \mathcal{S}_\star$.
\end{proposition}
Now, given the propagation of the constraints, Proposition \ref{Prop:PropagationofConstraints}, and Proposition \ref{Lemma:XCFEtoEFE} it follows
that the metric $\tilde{\bmg} = \Theta^{-2} \bmg$ obtained from the solution to the conformal evolution equations implies a
solution to the vacuum Einstein field equations with $\lambda=3$.

\medskip
The main result of this discussion is contained in the following theorem
\begin{theorem}
\label{Theorem:Main1}
Let $\hat{\mathbf{u}}_\star =\mathring{\mathbf{u}}_\star +\breve{\mathbf{u}}_\star$
denote smooth initial data for the conformal evolution equations satisfying
the conformal constraint equations on a hypersurface
$\mathcal{S}_\star$. Then, there exists $\varepsilon>0$ such that if
\[
||\breve{\mathbf{u}}_\star||_{\mathcal{S}_\star,m} < \varepsilon, \qquad
m\geq 4
\]
then there exists a unique $C^{m-2}$ solution $\tilde{\bmg}$ to the vacuum Einstein field equation with
positive Cosmological constant over $[\tilde{\tau}_\star,\infty)\times
\mathcal{S}_\star$ for $\tilde{\tau}_\star>0$ whose restriction to $\mathcal{S}_\star$ implies the
initial data $\hat{\mathbf{u}}_\star$. Moreover, the solution
$\hat{\mathbf{u}}$ remains suitably close to the background solution $\mathring{\bmu}$.
\end{theorem}

\section{Schwarzschild-de Sitter spacetimes}
In this section, it is discussed the behaviour of the conformal geodesics in the Cosmological region
of the sub-extremal Schwarzschild-de Sitter spacetime. The aim of this analysis is to adapt the technique described in the de Sitter-like setting and valid, in general, for asymptotically simple spacetimes to the black hole case as presented in \cite{MinVal23}.

\subsection{Basic properties}
The \emph{Schwarzschild-de Sitter spacetime} $(\tilde{\mathcal{M}},\mathring{\tilde{\bmg}})$ is a spherically symmetric solution to the 
  vacuum Einstein field equations with positive Cosmological constant \eqref{EFE} with $\tilde{\mathcal{M}}=\mathbb{R}\times\mathbb{R}^+\times
\mathbb{S}^2$ and line element given in \emph{standard coordinates}
$(t,r,\theta,\varphi)$ by
\begin{equation}	
\mathring{\tilde{\bmg}} =-\bigg{(} 1-\frac{2m}{r}-\frac{\lambda}{3}r^2\bigg{)}\mathbf{d} t  \otimes \mathbf{d} t +
\bigg{(}1-\frac{2m}{r}-\frac{\lambda}{3}r^2\bigg{)}^{-1} \mathbf{d} r \otimes  \mathbf{d} r + r^2  \bm{\sigma},
\end{equation}
where
\[
\bmsigma\equiv \mathbf{d}\theta\otimes\mathbf{d}\theta +\sin^2\theta \mathbf{d}\varphi\otimes\mathbf{d}\varphi,
\]
denotes the standard metric on $\mathbb{S}^2$. The coordinates
$(t,r,\theta,\varphi)$ take the range
\[
t\in(-\infty,\infty), \qquad r\in (0,\infty), \qquad
\theta\in (0,\pi), \qquad \varphi\in[0,2\pi).
\]
This line element can be rescaled so to that 
\begin{equation}	
\mathring{\tilde{\bmg}} =-D(r)\mathbf{d} t  \otimes \mathbf{d} t +
\frac{1}{D(r)} \mathbf{d} r \otimes  \mathbf{d} r + r^2  \bm{\sigma},
\label{BackgroundPhysicalMetric}
\end{equation}
where
\[
 D(r)\equiv 1 - \frac{M}{r} - r^2 \qquad {\rm and} \qquad M \equiv 2m \sqrt{\frac{ \lambda}{3}}.
\]
In our conventions $M$, $r$ and $\lambda$ are dimensionless quantities.

\subsection{Horizons and global structure}

The location of the horizons of the Schwarzschild-de Sitter spacetime
follows from the analysis of the zeros of the function $D(r)$ in the
line element \eqref{BackgroundPhysicalMetric}. 

\begin{figure}[t]
\centering
\includegraphics[width=1\textwidth]{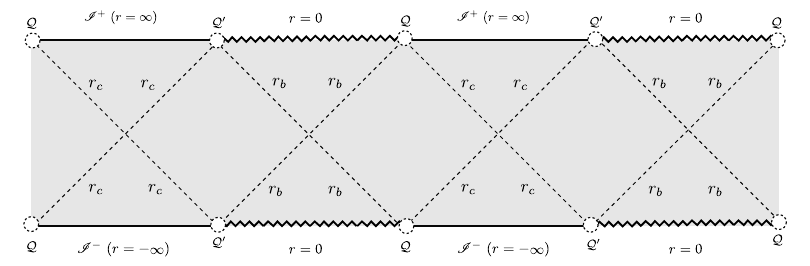}
\caption{Penrose diagram of the sub-extremal Schwarzschild-de Sitter
  spacetime. The serrated line denotes the location of the
  singularity; the continuous black line denotes the conformal
  boundary; the dashed line shows the location of the black hole and
  Cosmological horizons denoted by $\mathcal{H}_{b}$ and
  $\mathcal{H}_{c}$
  respectively. As described in the main text, these horizons are located
at $r=r_{b}$ and $r=r_{c}$.  The excluded points $\mathcal{Q}$
  and $\mathcal{Q'}$ where the singularity seems to meet the
  conformal boundary correspond to asymptotic regions of the
  spacetime that does not belong to the singularity nor the
  conformal boundary.  }
\label{Figure:SdSPenroseDiagram}
\end{figure}

\medskip
Since $\lambda>0$, the function $D(r)$ can be factorised as
\[
D(r) = -\frac{1}{r}(r-r_b)(r-r_c)(r-r_-), 
\]
where $r_b$ and $r_c$ are, in general, distinct positive roots of $D(r)$ and $r_-$ is
a negative root. Moreover, one has that 
\[
0<r_b<r_c, \qquad r_c+r_b+r_-=0.
\]
The root $r_b$ corresponds to a black hole-type of horizon and $r_c$
to a Cosmological de Sitter-like type of horizon. 
Using Cardano's formula for cubic equations, we have
\begin{subequations}
\begin{eqnarray}
&& r_-=- \frac{2}{\sqrt{3}} \cos \bigg{(}\frac{\phi}{3} \bigg{)}, \label{Horizon1}\\
&& r_b= \frac{1}{\sqrt{3}} \bigg{(} \cos \bigg{(}\frac{\phi}{3} \bigg{)}- \sqrt{3}  \sin \bigg{(}\frac{\phi}{3} \bigg{)} \bigg {)}, \label{Horizon2}\\
&& r_c= \frac{1}{\sqrt{3}} \bigg{(} \cos \bigg{(}\frac{\phi}{3}
   \bigg{)}+ \sqrt{3}  \sin \bigg{(}\frac{\phi}{3} \bigg{)} \bigg
   {)}. \label{Horizon3}
\end{eqnarray}
\end{subequations}
where the parameter $\phi$ is defined through the relation
\begin{equation}
  M= \frac{2 \cos \phi}{3 \sqrt{3}}, \qquad \phi \in \bigg{(} 0,
  \frac{\pi}{2} \bigg{)}.
  \label{DefinitionM}
\end{equation}
The sub-extremal case is characterised by $0 < M < 2/3 \sqrt{3}$
and $\phi \in (0, \pi/2)$ and describes a black hole in a
Cosmological setting. The Penrose diagram of the sub-extremal Schwarzschild-de Sitter is well
known ---see Figure \ref{Figure:SdSPenroseDiagram}. 

\subsection{Construction of a conformal Gaussian gauge in the
  Cosmological region}
\label{Section:CGSdS}
This study begins with the qualitative analysis of the behaviour of the conformal geodesics of the
Schwarzschild-de Sitter spacetime prescribed in terms of data
on hypersurfaces of constant $r$ in the Cosmological region.

\subsubsection{Basic setup}
It is assumed that
\[
r_c < r < \infty
\]
corresponding to the Cosmological region of the Schwarzschild-de
Sitter spacetime. Given a fixed $r=r_\star$, $\mathcal{S}_\star$ denotes the
spacelike hypersurface of constant $r=r_\star$ in this region. Points on
$\mathcal{S}_\star$ are described in terms of the coordinates
$(t,\theta,\varphi)$.

In order to prescribe the congruence of conformal geodesics, it is provided
the value of the conformal factor $\Theta_\star$
over $\mathcal{S}_\star$ so that
\[
  \Theta_\star=1, \qquad \dot{\Theta}_\star=0.
\]
The second condition implies that the resulting conformal factor will
have a time reflection symmetry with respect to
$\mathcal{S}_\star$. Then it is required that
\[
\tilde{\bmx}'_\star \perp \mathcal{S}_\star, \qquad \tilde{\beta}_\star=\Theta_\star^{-1}{\rm d}\Theta_\star.
\]
The latter, in turn, implies that 
\begin{equation}
\label{InitialData1} t=t_\star \qquad  t{}'{}_\star=
\frac{1}{\sqrt{D_\star}}, \qquad r{}'{}_\star=0, \qquad
\tilde{\beta}_{t \star}=0, \qquad\tilde{\beta}_{r \star}=0.
\end{equation}
These conditions give rise to a congruence of
conformal geodesics which has a trivial behaviour of the angular
coordinates. Accordingly,
the analysis of these curves is effectively given by the metric
\begin{equation}
\label{MetricWarped}
\tilde{\bmell} = 
      -D(r)\mathbf{d} t  \otimes \mathbf{d} t +
\frac{1}{D(r)} \mathbf{d} r \otimes  \mathbf{d} r.
  \end{equation}
Finally, in order to exclude the asymptotic points $\mathcal{Q}$ and $\mathcal{Q}'$, it is defined
\[
\mathcal{R}_\bullet =\{ p\in \mathcal{S}_\star \;|\; t(p) \in (-t_\bullet,t_\bullet) \},
\]
where the constant $t_\bullet$ is assumed large enough so that
$D^+(\mathcal{R}_\bullet)\cap \mathscr{I}^+\neq \varnothing$. 

\subsubsection{Analysis of the behaviour of the conformal geodesics}
The congruence of conformal geodesics prescribed by the initial data \eqref{InitialData1} is such that
$\beta^2=0$, so that after reparametrisation reduces to a congruence of metric geodesics. Thus, the geodesic equations imply that 
\begin{equation}
 r'=\sqrt{\gamma^2 - D(r)}, \qquad  D(r)t'{}^2-
 \frac{1}{D(r)}r'^2=1, \label{GeodesicEquations}
\end{equation}
where $\gamma$ is a constant. Evaluating at $\mathcal{S}_\star$ one
readily finds that
\[
t_\star'= \frac{|\gamma|}{|D_\star|},
\]
with $D_\star <0$. Moreover, since the unit normal to $\mathcal{S}_\star$ and $\tilde{\bmx}_\star'$ are
parallel to each other then $\gamma=0$.

\smallskip
In order to study the behaviour of these curves and obtain simpler expressions, it is set
$\lambda=3$ and $\tau_\star=0$. It follows then from Proposition \eqref{ConformalFactor}
that the conformal factor
\begin{equation}
  \Theta(\tau)=1-\frac{1}{4}\tau^2.
  \label{CanonicalThetaSdS}
  \end{equation}
Now, since the relation between the physical proper time $\tilde{\tau}$ and the unphysical proper time $\tau$ is obtained
from equation \eqref{CG:ChangeOfParameterFormula} so that 
\begin{equation}
  \tilde{\tau}= 2 {\rm arctanh}\bigg{(}\frac{\tau}{2} \bigg{)}, \qquad
  \tau=2 {\rm tanh}\bigg{(}\frac{\tilde{\tau}}{2} \bigg{)},
\label{TransformationProperTime}
\end{equation}
then
\[
\tau \rightarrow \pm 2, \qquad {\rm as} \hspace{0.5cm} \tilde{\tau} \rightarrow \pm \infty
\]
and since this congruence of conformal geodesics is reparametised as metric geodesics, it will reach the conformal boundary orthogonally ---see \cite{FriSch87}.
Now, since the dependence of the physical proper time $\tilde{\tau}$ on $r$ is given by
\[
\tilde{\tau}=\bigintss_{r_\star}^{r}\sqrt{
  \frac{\bar{r}}{(\bar{r}-r_b)(\bar{r}-r_c)(\bar{r}
    -r_-)}} {\rm d} \bar{r}
\]
which can be written in terms of elliptic functions ---see e.g. \cite{Law89}, it follows from the
general theory of elliptic functions that $\tilde{\tau}(r,r_\star)$ is
an analytic function of its arguments. Moreover, one has 
that 
\[
\tilde{\tau}\rightarrow \infty \qquad \mbox{as} \quad r\rightarrow \infty.
\]
Accordingly, the curves escape to infinity in an infinite
amount of physical proper time. Using the reparametrisation formulae
\eqref{TransformationProperTime} the latter corresponds to a finite
amount of unphysical proper time.

\subsubsection{Analysis of the behaviour of the conformal deviation equation}
In \cite{Fri03c} (see also \cite{GarGasVal18}) it has been shown that for congruences of conformal
geodesics in spherically symmetric spacetimes the behaviour of the
deviation vector of the congruence can be understood by considering
the evolution of a scalar $\tilde{\omega}$ satisfying the equation
\begin{equation}
\centernot{D}_{\tilde{x}'}\centernot{D}_{\tilde{x}'}\tilde{\omega}=\bigg{(}\bmbeta^2+\frac{1}{2}R[\tilde{\bmell}]\bigg{)}\tilde{\omega} + \centernot{D}_{\tilde{\bmz}}\bmbeta, \label{eveqomega}
\end{equation}
where $\centernot{D}$ denotes the Levi-Civita covariant derivative of $\tilde{\bmell}$ and $R[\tilde{\bmell}]$ denotes the Ricci scalar of $\tilde{\bmell}$.
If $\tilde{\omega}$ does not
vanish, then the congruence is non-intersecting. 

\begin{figure}[t]
\centering
\includegraphics[width=0.75\textwidth]{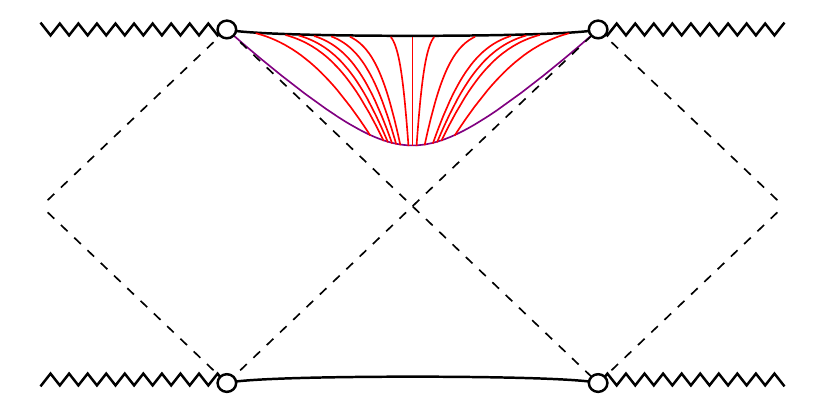}
\put(-230,153){$\mathcal{Q}$}
\put(-260,100){$r_b$}
\put(-210,100){$r_c$}
\put(-165,153){$\mathscr{I}^+$}
\put(-120,100){$r_c$}
\put(-65,100){$r_b$}
\put(-90,153){$\mathcal{Q}'$}
\put(-210,50){$r_c$}
\put(-260,50){$r_b$}
\put(-165,0.5){$\mathscr{I}^-$}
\put(-120,50){$r_c$}
\put(-65,50){$r_b$}
\put(-230,0.5){$\mathcal{Q}$}
\put(-90,0.5){$\mathcal{Q}'$}
\caption{The conformal geodesics are plotted on the Penrose diagram of
the Cosmological region of the sub-extremal Schwarzschild-de
Sitter spacetime. The purple line represents the initial hypersurface $\mathcal{S}_\star$ corresponding to $r=r_\star$. The red lines represent conformal geodesics with constant time leaving this initial hypersurface.}
\label{Figure:PenroseDiagramConformalGeodesics}
\end{figure}
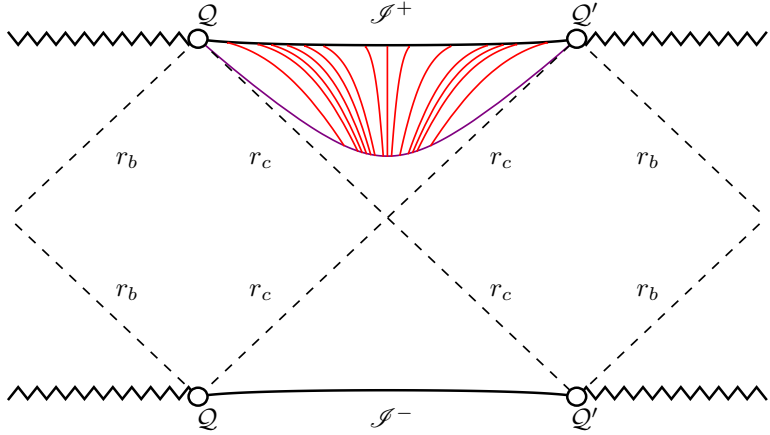

\smallskip
Since in the
present case one has $\bmbeta=0$ and $R[\tilde{\bmell}]=-\partial{}_r{}^2 D(r)$, it follows that the evolution equation
\eqref{eveqomega} takes the form
\[ 
\frac{{\rm d}^2 \tilde{\omega}}{{\rm d} \tilde{\tau}^2}=\bigg{(}1+
\frac{M}{r^3}\bigg{)}\tilde{\omega}, \qquad r\equiv r(\tilde{\tau},
r_\star).
\]
Since this setting $r\geq r_\star>r_c$ and $\omega \equiv \Theta \tilde{\omega}$, it follows that
\[ 
\frac{{\rm d}^2 \bar{\omega}}{{\rm d} \tilde{\tau}^2}=\bar{\omega},
\qquad \bar{\omega}(0,\rho_\star)= \frac{r_\star}{\rho_\star}, \qquad
\bar{\omega}'(0,\rho_\star)=0. 
\]
By solving this last differential equation and reverting to $\omega$, one has that
\[
 \omega \geq \frac{r_\star}{\rho_\star} \bigg{(} 1+ \frac{
   \tau^2}{4} \bigg{)} >0,
\]
which is non-vanishing in the limit $\tau \rightarrow \pm 2$.
Thus, we have the following Proposition
\begin{proposition}
\label{Proposition:CGNoCaustics}
The congruence of conformal geodesics given by the initial conditions
\eqref{InitialData1} leaving the initial hypersurface
$\mathcal{S}_\star$ reach the conformal boundary $\mathscr{I}^+$
without developing caustics.
\end{proposition}
This behaviour of the conformal geodesics is shown in Figure [$\ref{Figure:PenroseDiagramConformalGeodesics}$].

\subsubsection{Conformal Gaussian coordinates in the sub-extremal
  Schwarzschild-de Sitter spacetime}
\label{Subsection:CGCoordinates}
The congruence of conformal geodesics defined by the initial
conditions \eqref{InitialData1} is used to construct a
\emph{conformal Gaussian coordinate system} in a domain in the chronological
future of $\mathcal{R}_\bullet$ containing a
portion of the conformal boundary $\mathscr{I}^+$. This analysis is carried out by considering the coordinate $z\equiv 1/r$ in terms of which the line element
\eqref{BackgroundPhysicalMetric} takes the form
\[
  \mathring{\tilde{\bmg}} =\frac{1}{z^2} \bigg{(} -F(z)\mathbf{d} t  \otimes \mathbf{d} t +
\frac{1}{F(z)} \mathbf{d} z \otimes  \mathbf{d} z +
\bm{\sigma}\bigg{)},
\]
where
\[
  F(z)\equiv  z^2 D(1/z).
\]
The above expression suggest defining an \emph{unphysical metric} $\bar{\bmg}$ via
\[
  \bar{\bmg}= \Xi^2 \mathring{\tilde{\bmg}}, \qquad \Xi \equiv  z.
\]
More precisely, one has
\begin{equation}
  \bar{\bmg}=- F(z)\mathbf{d} t  \otimes \mathbf{d} t +
  \frac{1}{F(z)} \mathbf{d} z \otimes  \mathbf{d} z + \bm{\sigma}.
  \label{AuxiliaryConformalMetric}
\end{equation}
Now, let $\widetilde{SdS}_I$ denote the Cosmological region of the
Schwarzschild-de Sitter spacetime ---that is
\[
\widetilde{SdS}_I =\{ p\in \tilde{\mathcal{M}} \; |\; r(p)>r_c \}.
\]
Moreover, denote by  $SdS_I$ the conformal representation of
$\widetilde{SdS}_I$ defined by the conformal factor $\Theta$ defined by
the non-singular congruence of conformal geodesics. Let $z\equiv
1/r$, for $z < z_c$ one has that in terms of these coordinates
\begin{equation}
SdS_I = \{ p\in \mathbb{R}\times \mathbb{R}\times \mathbb{S}^2 \; |\;
0\leq z(p) \leq z_\star\},
\label{Definition:SdSI}
\end{equation}
where $z_\star\equiv 1/r_\star$ with $r_\star > r_c$. 

\medskip
The conformal
geodesics defined by the initial conditions \eqref{InitialData1} define a map $\psi$ which is analytic in the parameters $(\tau,t_\star)$. This map is invertible since the Jacobian of the transformation is
non-zero for the given value of the parameters. The inverse map 
\[
\psi^{-1}:  [0,z_\star]\times [-t_\bullet, t_\bullet]
\rightarrow [0,2] \times [-t_\bullet, t_\bullet], \qquad  (t,z) \mapsto \big(\tau(t,z),t_\star(t,z) \big)
\]
 gives the transformation from the
\emph{standard Schwarzschild
coordinates} $(t,z,\theta,\varphi)$ into the \emph{conformal Gaussian
coordinates} $(\tau, t_\star, \theta,\varphi)$. This result is summarised by the following
\begin{proposition}
  \label{Proposition:CGCoordinates}
The congruence of conformal geodesics on $SdS_I$ defined by the
initial conditions on $\mathcal{S}_\star$ given by
\eqref{InitialData1} induce a conformal Gaussian coordinate system
over $D^+(\mathcal{R}_\bullet)$ which is related to the standard coordinates
$(t,r)$ via a map which is analytic. 
\end{proposition}

\subsection{The background spacetime as a solution to the conformal Einstein field equations}
\label{Section:SdSGaussian}
The Schwarzschild-de Sitter spacetime in the region 
\[
\mathcal{M}_\bullet \equiv [0,2] \times [-t_\bullet, t_\bullet]
\]
is casted as a solution to the extended conformal Einstein field equations by means of a Weyl
  propagated frame. 

\subsubsection{The frame}
  Since the congruence of conformal geodesics implied by the initial data
  \eqref{InitialData1} satisfies $\tilde{\bmbeta}=0$, the Weyl propagation equation
  \eqref{WeylPropagation}  reduces to the usual parallel
  propagation equation. Given the spherical symmetry of the Schwarzschild-de Sitter
spacetime, the discussion of a frame adapted to the symmetry of the
spacetime can be carried out by considering the 2-dimensional
Lorentzian metric \eqref{MetricWarped}.
The \emph{time leg} of the frame is set as $\bme_\bmzero=\dot{\bmx}$ so that
\[
  \bme_\bmzero=\Theta^{-1} \tilde{\bmx}',
\]
where $\tilde{\bmx}' = \tilde{t}'\bmpartial_t + \tilde{r}'\bmpartial_r $. Now, upon defining
\[
\bmomega\equiv \epsilon_\bmell(\tilde{\bmx}', \cdot )
\]
with $\langle \bmomega, \tilde{\bmx}'\rangle =0$, one has that the \emph{radial leg} of the frame is given by
\[
\bme_\bmone =\Theta \bmomega^\sharp.
\]
The Weyl propagated frame $\{ \bme_\bma\}$ is completed by choosing \emph{two arbitrary orthonormal
vectors} $\tilde{\bme}_{\bmtwo\star}$ and $\tilde{\bme}_{\bmthree\star}$ spanning the
tangent space of $\mathbb{S}^2$ and defining the vectors $\{ \bme_\bmtwo,
\bme_\bmthree \}$ on $\mathcal{M}_\bullet$ by constantly extending 
the value of the associated 
coefficients along the conformal
geodesics. This analysis leads to the following result

\begin{proposition}
  \label{Proposition:SdSWeylPropagatedFrame}
Let $\tilde{\bmx}'$ denote the vector tangent to the conformal
geodesics defined by the initial data \eqref{InitialData1} and let $\{
\bme_{\bmtwo\star},\, \bme_{\bmthree\star} \}$ be an arbitrary
orthonormal pair of vectors spanning the tangent bundle of
$\mathbb{S}^2$.
Then the frame $\{\bme_\bmzero,\,
\bme_\bmone,\,\bme_\bmtwo,\,\bme_\bmthree\}$ obtained by the procedure
 described in the previous paragraph is a $\bmg$-orthonormal Weyl
propagated frame. The frame depends analytically on the unphysical
proper time $\tau$ and the initial position $t_\star$ of the curve. 
\end{proposition}

  \subsubsection{The Weyl connection}
  \label{Subsection:WeylConnection}
 The connection coefficients associated to a conformal Gaussian gauge
 are made up of two pieces: the 1-form defining the Weyl connection
 and the Levi-Civita connection of the metric $\bar{\bmg}$. 

\smallskip
The congruence of conformal geodesics discussed in Section \ref{Section:CGSdS} arises from initial data chosen so that the curves with tangent given
by $\tilde{\bmx}'$ satisfy the standard (affine) geodesic
equation. Consequently, the (spatial) 1-form
$\tilde{\bmbeta}$ vanishes. Now, since
  $\tilde{\bmx}' = r' \bmpartial_r$, by observing equation 
  \eqref{GeodesicEquations} for $r'$ with $\gamma =0$ and by introducing $z=1/r$, it follows
that 
\[
\bmbeta \approx -\frac{1}{z}\mathbf{d}z \qquad \mbox{for}\quad
z\approx 0.
\]
Then, from
the conformal transformation rule 
\[
\bar{\bmbeta} = \bmbeta + \Xi^{-1} \mathbf{d}\Xi 
\]
 and by recalling that $\Xi=z$, it follows that $\bar{\bmbeta}$
vanishes at $\mathscr{I}^+$. However,
$\bar{\bmbeta}\neq 0$ away from the conformal boundary.

\subsubsection{The connection coefficients}
Since the coordinates and
connection coefficients associated to the physical connection
$\tilde{\bmnabla}$ are not well adapted to a discussion near the
conformal boundary we resort to the unphysical Levi-Civita connection
$\bar{\bmnabla}$ to compute $\hat{\bmnabla}$. 

\medskip
The connection
coefficients $\hat{\Gamma}_\bma{}^\bmb{}_\bmc$ are defined through the
relation
\[
\hat{\nabla}_\bma \bme_\bmc = \hat{\Gamma}_\bma{}^\bmb{}_\bmc \bme_\bmb.
\]
The only non-vanishing
Christoffel symbols 
$\bar{\Gamma}_\mu{}^\nu{}_\lambda$ are given by
\begin{eqnarray*}
&&\bar{\Gamma}_t{}^t{}_z = -\bar{\Gamma}_z{}^z{}_z = \frac{z
  (\frac{3}{2}Mz-1)}{1 + z^2(Mz-1)}, \\
&&  \bar{\Gamma}_t{}^z{}_t = z(\tfrac{3}{2}Mz-1)\big( 1
  +z^2(Mz-1) \big),\\
&& \bar{\Gamma}_\varphi{}^\theta{}_\varphi =-\cos\theta \sin\theta, \qquad
   \bar{\Gamma}_\theta{}^\varphi{}_\varphi=\cot \theta.  
  \end{eqnarray*}
These coefficients are analytic at $z=0$.
Since a contraction with the coefficients of the frame does not change
this, it follows that the Weyl connection coefficients
$\hat{\Gamma}_\bma{}^\bmb{}_\bmc$ are smooth functions of the
coordinates used in the conformal Gaussian gauge on the future of the
fiduciary initial hypersurface $\mathcal{S}_\star$ up to and beyond
the conformal boundary.

\subsubsection{The rescaled Weyl tensor}
Given a timelike vector, the components of the rescaled Weyl tensor
$d_{abcd}$ can be encoded in the electric and magnetic parts relative
to the given vector. For the vector $ \bar{\bme}_\bmzero$ these are
given by
\[
 d_{ac}=d_{abcd}\bar{e}_\bmzero{}^b \bar{e}_\bmzero {}^{d}, \qquad d^{*}{}_{ac}=d{}^*{}_{abcd}\bar{e}_{\bmzero}{}^{b}\bar{e}_\bmzero{}^{d},
\]
where $d{}^*{}_{abcd}$ denotes the Hodge dual of $d_{abcd}$. A
computation using the package {\tt xAct} for {\tt Mathematica} readily
gives that the only non-zero components of the electric part are given by
\begin{eqnarray*}
&&d_{tt}= -M\big{(} z^2(1-Mz)-1 \big{)}, \\
&& d_{\theta\theta}=-\frac{M}{2},\\
&& d_{\varphi\varphi}=-\frac{M}{2}\sin^2\theta,  
\end{eqnarray*}
while the magnetic part vanishes identically. These expressions are regular at $z=0$ --- by disregarding
the coordinate singularity due to the use of spherical
coordinates. The smoothness of the components of the Weyl tensor is
retained when contracted with the coefficients of the frame $\{\bme_\bma\}$.

\subsubsection{The Schouten tensor}
A similar computer algebra calculation shows that the non-zero
components of the Schouten tensor of the metric $\bar{\bmg}$ are given by
\begin{eqnarray*}
&&\bar{L}_{tt}=\frac{1}{2} (2Mz-1)(1 +  z^2(Mz-1)),\\
&&\bar{L}_{zz}=-\frac{1}{2}\frac{(2Mz-1)}{1 +z^2(Mz-1)},\\
&&\bar{L}_{\theta\theta}=-\frac{1}{2}(Mz-1),\\
  && \bar{L}_{\varphi\varphi}=-\frac{1}{2} \sin^2\theta(Mz-1).
\end{eqnarray*}
The above expressions are analytic on
$\mathcal{M}_\bullet$ ---in particular at $z=0$ and by disregarding the coordinate singularity on the angular
components. To obtain the
components of the Schouten tensor associated to the Weyl connection
$\hat{\bmnabla}$ we make use of the transformation rule
\[
\bar{L}_{ab}-\hat{L}_{ab} = \bar{\nabla}_a \bar{\beta}_b - \frac{1}{2}S_{ab}{}^{cd}\bar{\beta}_c\bar{\beta}_d.
\]
The smoothness of $\bar{\beta}_a$ has already been established in
Subsection \ref{Subsection:WeylConnection}. Thus, the
components of $\hat{L}_{ab}$ with respect to the Weyl propagated frame
$\{\bme_\bma\}$ are regular on $\mathcal{M}_\bullet$.

\subsubsection{Construction of a background solution with compact spatial
  sections}
From the previous discussion, it follows that the sub-extremal Schwarzschild-de Sitter spacetime
expressed in terms of a conformal Gaussian gauge system gives rise to
a solution to the extended conformal Einstein field equations on the
region $\mathcal{M}_\bullet \subset D^+(\mathcal{R}_\bullet)$.
Since $\mathcal{R}_\bullet\subset \mathcal{S}_\star$ has the
topology of $I\times \mathbb{S}^2$ where $I\subset \mathbb{R}$ is an
open interval, the spacetime arising from
$\mathcal{R}_\bullet$ will have spatial sections with the same
topology. As part of the perturbative argument is
based on the general theory of symmetric hyperbolic systems as given
in \cite{Kat75b} it is convenient to consider solutions with compact
spatial sections.

\begin{figure}[t]
\centering
\includegraphics[width=0.9\textwidth]{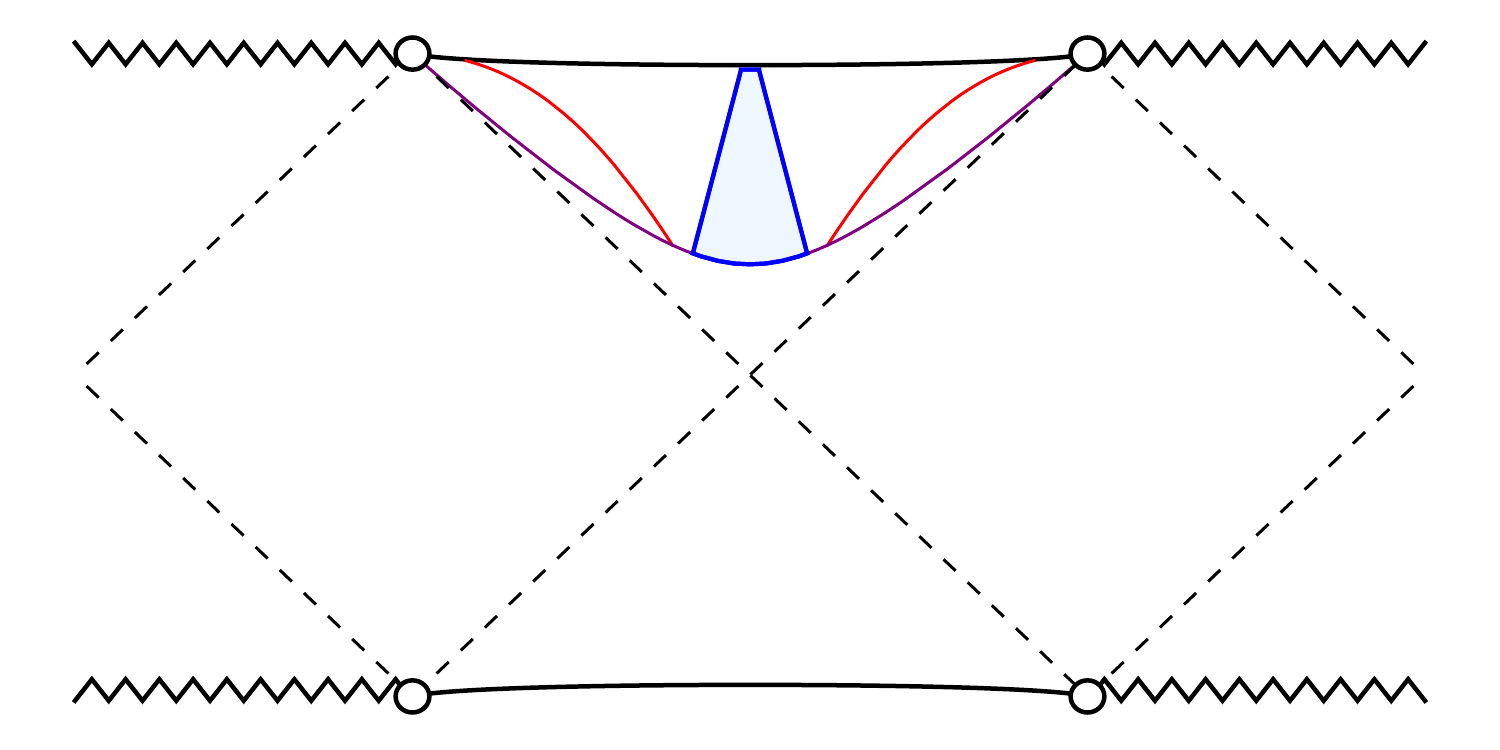}
\put(-280,185){$\mathcal{Q}$}
\put(-310,130){$r_b$}
\put(-245,130){$r_c$}
\put(-195,185){$\mathscr{I}^+$}
\put(-215,118){$-t_\bullet$}
\put(-140,130){$r_c$}
\put(-75,130){$r_b$}
\put(-110,185){$\mathcal{Q}'$}
\put(-245,60){$r_c$}
\put(-310,60){$r_b$}
\put(-195,1){$\mathscr{I}^-$}
\put(-145,60){$r_c$}
\put(-75,60){$r_b$}
\put(-280,1){$\mathcal{Q}$}
\put(-110,1){$\mathcal{Q}'$}
\put(-180,118){$t_\bullet$}
\put(-230,160){$\mathcal{T}_{-2t_\bullet}$}
\put(-170,160){$\mathcal{T}_{2t_\bullet}$}
\caption{The red curves identify the timelike hypersurfaces
  $\mathcal{T}_{-2t_\bullet}$ and $\mathcal{T}_{2t_\bullet}$. The resulting spacetime manifold $\bar{M}_\bullet$ has compact spatial sections, $\bar{\mathcal{S}}_z$,with the topology of $\mathbb{S}^1\times\mathbb{S}^2$.}
\label{Figure:FutureDomainofDependenceWithExtension}
\end{figure}
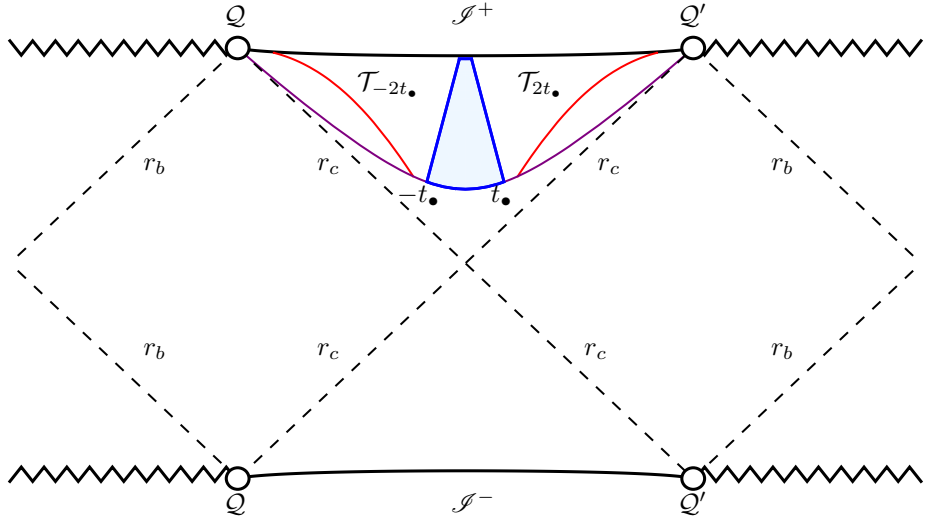

\smallskip
This construction is based on the observation that the Killing vector
$\bmxi=\bmpartial_t$ in the Cosmological region of the spacetime is
spacelike. Thus, given a fixed $z_\circ<z_c$, the
hypersurface $\mathcal{S}_{z_\circ}$ defined by the condition
$z=z_\circ$ has a translational invariance. Now, one identifies
the timelike hypersurfaces $\mathcal{T}_{-2t_\bullet}$
and $\mathcal{T}_{2t_\bullet}$ generated,
respectively, by the future-directed geodesics emanating from
$\mathcal{S}_\star$ at the points with $t=-2t_\bullet$ and
$t=2t_\bullet$ to
obtain a smooth spacetime manifold $\bar{\mathcal{M}}_\bullet$ with
compact spatial sections ---see Figure
\ref{Figure:FutureDomainofDependenceWithExtension}. 
The metric $\bar{\bmg}$ on $SdS_I$ induces a metric on
$\bar{\mathcal{M}}_\bullet$ which, on an abuse of notation, is denoted
again by $\bar{\bmg}$. Since the initial conditions defining the
congruence of conformal geodesics of Section \ref{Section:CGSdS} have translational invariance, the resulting curves also have this property. Accordingly,
the congruence of conformal geodesics on $SdS_I$ induces a non-intersecting congruence
of conformal geodesics on $\bar{\mathcal{M}}_\bullet$. Thus, the solution to the extended conformal Einstein field
equations in a conformal Gaussian gauge implies a similar
solution over the manifold $\bar{\mathcal{M}}_\bullet$ denoted by $\mathring{\mathbf{u}}$. The
initial data induced by $\mathring{\mathbf{u}}$ on
$\bar{\mathcal{S}}_\star$ will be denoted by
$\mathring{\mathbf{u}}_\star$.

\subsection{Structural properties of the evolution and subsidiary equations}
\label{Section:EvolutionEqns}
The conformal Gaussian gauge system leads to a \emph{hyperbolic reduction}
of the extended conformal Einstein field equation \eqref{ecfe5}. The
particular form of the resulting evolution equations is not 
required in this analysis, only general structural properties. 

\medskip
The extended conformal Einstein field equations \eqref{ecfe5}
expressed in in terms of a conformal Gaussian gauge imply evolution equations in the form of a symmetric
hyperbolic system for the components ${\bmupsilon}\equiv  (\bme, {\bmGamma}, {\bmL})$ and ${\bmphi}\equiv  ({\bmd}, {\bmd}^*)$ as in Lemma \ref{Lemma:EvolutionEqns}.
Now, since the evolution equations hold. Then the independent
components of the zero-quantities
\[
{\Sigma}_\bma{}^\bmb{}_\bmc, \quad
{\Xi}^\bmc{}_{\bmd\bma\bmb}, \quad {\Delta}_{\bma\bmb\bmc},
\quad \Lambda_{\bma\bmb\bmc}, \quad \delta_\bma,\quad
\gamma_{\bma\bmb}, \quad \varsigma_{\bma\bmb},
\]
not determined by either the evolution equations or the gauge
conditions satisfy a symmetric hyperbolic system which is homogeneous
in the zero-quantities. As a result, if the zero-quantities vanish on
a fiduciary spacelike hypersurface $\mathcal{\bar{S}}_\star$, then they also
vanish on the domain of dependence ---see \cite{Kat75b}.

\subsection{The perturbative existence argument}
Let $\hat{\mathbf{u}} \equiv (\hat{\bmv},\hat{\bmphi})$ and $\mathring{\mathbf{u}}$ denotes the \emph{background
  solution} being a solution to the evolution
equations arising from the initial data
$\mathring{\mathbf{u}}_\star$ prescribed on $\bar{\mathcal{S}}_\star$. Solutions to the evolution equations which can be regarded as a perturbation of the
background solution are constructed by introducing a perturbative argument
\[
\hat{\mathbf{u}}= \mathring{\mathbf{u}} + \breve{\mathbf{u}}
\]
with $\breve{\mathbf{u}}$ being a small perturbation.
This means, in particular, that one can write
\begin{equation}
  \label{split}
  \hat{\bme}=\mathring{\bme} + \breve{\bme}, \qquad \hat{\bmGamma}=\mathring{\bmGamma} + \breve{\bmGamma}, \qquad \hat{\bmphi}=\mathring{\bmphi} + \breve{\bmphi}.
\end{equation}
The components of $\breve{\bme}$, $\breve{\bmGamma}$ and
$\breve{\bmphi}$ are our unknowns. Making use of the decomposition
\eqref{split} and exploiting that  $\mathring{\bmu}$ is a solution to
the conformal evolution equations one obtains the equations 
\begin{subequations}
\begin{align}
& \label{he3}\partial_\tau \breve{\bmupsilon}= \mathbf{K} \breve{\bmupsilon} + \mathbf{Q}(\mathring{\bmGamma}+ \breve{\bmGamma}) \breve{\bmupsilon}+ \mathbf{Q}(\breve{\bmGamma})\mathring{\bmupsilon} + \mathbf{L}(\bar{x}) \breve{\bmphi}+ \mathbf{L}(\bar{x}) \mathring{\bmphi},  \\
&\label{he4} (\mathbf{I} + \mathbf{A}^0(\mathring{\bme} + \breve{\bme})) \partial_\tau \breve{\bmphi}+ \mathbf{A}^\alpha(\mathring{\bme} + \breve{\bme}) \partial_\alpha \breve{\bmphi}=\mathbf{B}(\mathring{\bmGamma}+ \breve{\bmGamma}) \breve{\bmphi}+\mathbf{B}(\mathring{\bmGamma}+ \breve{\bmGamma}) \mathring{\bmphi}.
\end{align}
\end{subequations}
Now, it is convenient to define
\[
\bar{\mathbf{A}}^0(\tau, \underline{x}, \breve{\mathbf{u}}) \equiv \begin{pmatrix*}
      \mathbf{I} &  0 \\
      0 & \mathbf{I} + \mathbf{A}^0(\mathring{\bme}+ \breve{\bme})
     \end{pmatrix*},
\qquad 
\bar{\mathbf{A}}^\alpha(\tau, \underline{x}, \breve{\mathbf{u}}) \equiv \begin{pmatrix*}
  0 &   0 \\
      0 &   \mathbf{A}^\alpha(\mathring{\bme}+ \breve{\bme})
     \end{pmatrix*},
   \]
  and
\[
\bar{\mathbf{B}}(\tau, \underline{x}, \breve{\mathbf{u}})\equiv \breve{\mathbf{u}}\bar{\mathbf{Q}}\breve{\mathbf{u}}+ \bar{\mathbf{L}}(\bar{x})\breve{\mathbf{u}}+ \bar{\mathbf{K}}\breve{\mathbf{u}},
\] 
where 
\[
\breve{\mathbf{u}}\bar{\mathbf{Q}}\breve{\mathbf{u}} \equiv \begin{pmatrix*}
    \breve{\bmupsilon}\mathbf{Q}\breve{\bmupsilon} &     0 \\
     0 &
     \mathbf{B}(\breve{\bmGamma})\breve{\bmphi} +  \mathbf{B}(\breve{\bmGamma})\mathring{\bmphi}
     \end{pmatrix*},
\qquad 
\bar{\mathbf{L}}(\bar{x})\breve{\mathbf{u}} \equiv \begin{pmatrix*}
  \mathring{\bmupsilon}\mathbf{Q}\breve{\bmupsilon}+ \mathbf{Q}(\breve{\bmGamma})\mathring{\bmupsilon}&
    \mathbf{L}(\bar{x})\breve{\bmphi}+ \mathbf{L}(\bar{x})\mathring{\bmphi}  \\
      0 &  0
     \end{pmatrix*},
\]
\[
\bar{\mathbf{K}} \breve{\mathbf{u}}\equiv \begin{pmatrix*}
 \mathbf{K} \breve{\bmupsilon} &  0  \\
      0 & \mathbf{B}(\mathring{\Gamma})\breve{\bmphi}+ \mathbf{B}(\mathring{\Gamma})\mathring{\bmphi}
     \end{pmatrix*},
\]
denote, respectively, expressions which are quadratic, linear and constant terms in the unknowns. 

\medskip
In terms of the above expressions it is possible to rewrite the system
\eqref{he3}-\eqref{he4} in the more concise form 
\begin{equation}
 \bar{\mathbf{A}}^0(\tau, \underline{x}, \breve{\mathbf{u}})\partial_\tau \breve{\mathbf{u}}+ \bar{\mathbf{A}}^\alpha(\tau, \underline{x}, \breve{\mathbf{u}})\partial_\alpha \breve{\mathbf{u}}=\bar{\mathbf{B}}(\tau, \underline{x}, \breve{\mathbf{u}}). \label{he5}
\end{equation}
These equations are in a form where the theory of first-order
symmetric hyperbolic systems \cite{Kat75c} can be applied to obtain a existence and
stability result for small perturbations of the initial data
$\mathring{\mathbf{u}}_\star$. 

\subsection{A solution to the Einstein field equations}
The evolution equations \eqref{he3}-\eqref{he4} imply the same subsidiary system as for de Sitter-like spacetimes. Thus, the \emph{propagation of the constraints} follows from the same argument ---see Proposition 3.1 Section \ref{Section:PropagationofConstraints}. Given the propagation of the constraints and Proposition \ref{Lemma:XCFEtoEFE}, one has the metric $\tilde{\bmg} = \Theta^{-2} \bmg$
obtained from the solution to the conformal evolution equations implies a
solution to the vacuum Einstein field equations with positive
Cosmological constant on $\tilde{\mathcal{M}}\equiv D^+(\mathcal{R}_\bullet)$. 

\medskip
The main result of this discussion is contained in the following theorem
\begin{theorem}
\label{Theorem:MainSdS2}
Let $\hat{\mathbf{u}}_\star =\mathring{\mathbf{u}}_\star +\breve{\mathbf{u}}_\star$
denote smooth initial data for the conformal evolution equations satisfying
the conformal constraint equations on a hypersurface
$\overline{\mathcal{S}}_\star$. Then, there exists $\varepsilon>0$ such that if
\[
||\breve{\mathbf{u}}_\star||_{\overline{\mathcal{S}}_\star,m} < \varepsilon, \qquad
m\geq 4
\]
then there exists a unique $C^{m-2}$ solution $\tilde{\bmg}$ to the vacuum Einstein field equation with
positive Cosmological constant over $[\tilde{\tau}_\star,\infty)\times
\overline{\mathcal{S}}_\star$ for $\tilde{\tau}_\star>0$ whose restriction to $\overline{\mathcal{S}}_\star$ implies the
initial data $\hat{\mathbf{u}}_\star$. Moreover, the solution
$\hat{\mathbf{u}}$ remains suitably close to the background solution $\mathring{\mathbf{u}}$.
\end{theorem}
In particular, the resulting spacetime $(\tilde{\mathcal{M}},\tilde{\bmg})$ is a non-linear perturbation of the sub-extremal Schwarzschild-de Sitter spacetime on a portion of the Cosmological region of the
  background solution which contains a portion of the asymptotic region.

\section{Conclusion}
This review article provides a discussion based on \cite{MinVal21} and \cite{MinVal23} describing how the extended conformal Einstein field equations and a gauge adapted to the conformal geodesics can be used to study the evolution of vacuum spacetimes with positive Cosmological constant. In the Sitter-like case, this analysis identifies a class of spacetimes for which it is possible to prove non-linear stability and the existence of a regular conformal
representation. More precisely, it is identified a class of de Sitter-like spacetimes which can be conformally embedded
into a portion of a cylinder whose sections have negative scalar
curvature. The conformal embedding is realised by means of a conformal
factor $\Theta$ which depends quadratically on the affine parameter
$\tau$ of the conformal geodesics and
this parameter is also used as a time coordinate for the physical
metric. This result led to wonder whether this technique can be adapted to black hole type of spacetimes. The analysis of the conformal geodesics in the Cosmological region of the Schwarzschild-de Sitter spacetime shows that it is possible to construct a conformal Gaussian gauge system. In particular, it shows that it is possible to construct solutions to the vacuum Einstein field equations in this region containing a portion of
the asymptotic region and which are non-linear
perturbations of the exact Schwarzschild-de Sitter
spacetime. Crucially, although the spacetimes constructed have an
infinite extent to the future, they exclude the \emph{asymptotic points} $\mathcal{Q}$ and $\mathcal{Q}'$. From the analysis of the asymptotic initial value problem
in \cite{GasVal17a} it is known that these points contain
singularities of the conformal structure. Thus, they cannot be dealt
by the approach used in the article. In order to have a complete statement on the non-linear stability of the Cosmological region it is necessary to address the asymptotic points. Moreover, since the initial hypersurfaces $\mathcal{S}_\star$ considered in the article are spacelike and the evolution doesn't include the Cosmological horizon $r_c$. A complete statement should also include the case in which $r=r_c$. This suggests reformulating the existence and
stability results in \cite{MinVal23} in terms of a characteristic initial
value problem with data prescribed on Cosmological horizons. Again,
to avoid the singularities of the conformal structure, the
characteristic data has to be prescribed away from the asymptotic points. Alternatively, one
could consider data sets which become exactly Schwarzschild-de Sitter
near the asymptotic points. The associated evolution problem by means of a generalisation of the methods used in
\cite{HilValZha20b} should allow to reach any suitable hypersurface with constant r.


 \end{document}